\documentclass[twocolumn,apj]{emulateapj}

\def\mbf#1{\mbox{\boldmath ${#1}$}}
\def\Alfven{Alfv\'{e}n~}
\def\Alfvenic{Alfv\'{e}nic~}

\slugcomment{Submitted to ApJ}

\shorttitle{Structured Red Giant Winds and the Dividing Line}
\shortauthors{T. K. Suzuki}

\begin{document}

\title{Structured Red Giant Winds with Magnetized Hot Bubbles and 
the Corona/Cool Wind Dividing Line}

\author{Takeru K. Suzuki$^{1}$}
\email{stakeru@ea.c.u-tokyo.ac.jp}
\altaffiltext{1}{School of Arts and Sciences, University of Tokyo,
Komaba, Meguro, Tokyo, Japan 153-8902}

\begin{abstract}
By performing magnetohydrodynamical (MHD) simulations, we investigate 
the mass loss of intermediate- and low-mass stars 
from main sequence (MS) to red giant branch (RGB) phases. 
\Alfven waves, which are excited by 
the surface convections travel outwardly and dissipate 
by nonlinear processes to accelerate and heat the stellar winds. 
We dynamically treat these processes in open magnetic field regions from the 
photospheres to $\simeq$25 stellar radii. 
When the stars evolve to slightly blueward positions of the dividing line 
(Linsky \& Haisch), the steady hot corona with temperature, 
$T \approx 10^6$ K, suddenly disappears. Instead, many hot ($\sim 10^6$ K) 
and warm ($\gtrsim 10^5$ K) bubbles are formed in cool 
($T\lesssim 2\times 10^4$ K) chromospheric winds because of 
thermal instability; 
the red giant wind is not a steady stream but structured outflow.
As a result, the mass loss rates, $\dot{M}$, largely vary in time by 3-4 
orders or magnitude in the RGB stars. 
Supported by magnetic pressure, the density of hot bubbles  
can be kept low to reduce the radiative cooling and to maintain the high 
temperature long time.     
Even in the stars redward of the dividing line, 
hot bubbles intermittently exist, and 
they can be sources of ultraviolet/soft X-ray emissions from hybrid stars.  
Nearly static regions are formed above the photospheres of the RGB stars, and 
the stellar winds are effectively accelerated from several stellar radii. 
Then, the wind velocity is much smaller than the surface escape speed, 
because it is regulated by the slower escape speed at that location. 
We finally derive an equation that determines $\dot{M}$ from the
energetics of the simulated wave-driven winds in a {\it forward} manner.     
The relation explains $\dot{M}$ from MS to RGB, and it 
can play a complementary role to the Reimers' formula, which is mainly for 
more luminous gaints.

\end{abstract}

\section{Introduction}
Solar-type stars have X-ray emitting coronae with the 
temperatures, $T\gtrsim 10^6$ K, and hot stellar winds stream out 
with the asymptotic terminal velocities, $v_{\infty}=100-1000$ km s$^{-1}$, 
comparable to the escape velocities at the surfaces. 
The total mass lost by the wind is negligible during the main sequence 
(MS) phase in comparison with the stellar mass, whereas the wind may play 
a role in the loss of the angular momentums. 
Along with the stellar evolution to red giant branch (RGB), 
the mass loss rate, $\dot{M}$'s, increases by 4-10 orders of 
magnitude \citep{js91}, while 
$v_{\infty}$ drops to $10-100$ km s$^{-1}$, which is smaller than the surface 
escape velocity.  In addition, the temperature of the wind also decreases, 
which is probably drastic across the ``dividing line'' (DL hereafter; 
Linsky \& Haisch 1979) discussed below. 
The mass loss during the RGB phase becomes more important, and it may control 
the later evolution of a star itself. 
For instance, some stars evolve to a peculiar group of blue horizontal 
branch stars, and it is inferred that it is related to the mass loss 
during the RGB phase \citep{ydk97}.      
In spite of its importance, however, the acceleration mechanism of red giant 
winds is not well understood.



It was reported that cool single RGB stars redward of the DL 
(near spectral type K1 III in Hertzsprung-Russel (HR)-diagram;  
Figure \ref{fig:hr}) showed an evidence of neither ultraviolet (UV) radiation 
of C$_{\rm IV}$ ($T\sim 10^5$ K) \citep{lh79} nor soft X-ray ($T\sim 10^6$ K) 
\citep{ayr81}. 
\citet{sm80} further pointed out that the stellar winds appear cool and 
massive after stars cross the DL.   
However, \citet{hdr80} observed late-type super-giants in the 
noncoronal zone that exhibit 
both UV emission and blue-shifted chromospheric absorption features, 
indicating that hot gas and cool outflow coexist. 
Such hybrid stars were also detected 
by later observations (e.g. Ayres et al. 1998), whereas the activity levels 
of UV/soft X-ray are much lower than in the coronal zone. 
These observations show that the simple DL picture roughly holds 
while the reality seems a bit more complicated.

RGB stars are generally slow rotators. The radiation luminosities of 
intermediate- and low-mass stars near the DL are not 
still large, either. 
Therefore, the centrifugal force and radiation pressure are not 
sufficient in accelerating the stellar winds (e.g. Judge \& Stencel 1991). 
Instead, the surface convective layer is expected to be 
the main origin of driving the winds and UV/X-ray activities. 
Magnetic fields are probably generated by turbulent dynamo in the 
convection zone. 
The energy of the surface turbulences is lifted up through the magnetic 
fields, and the dissipation of the energy in upper regions leads to the 
heating of the atmosphere and the acceleration of the wind. 
This picture is essentially the same as what takes place in the solar corona 
and wind. Difference from the sun is that the surface gravity is much 
smaller in RGB stars. In a na\"{i}ve way of thinking, the transition from 
the steady coronae to the cool winds (with some hybrid activities) is possibly 
explained by this gravity effect even though other conditions are similar; 
the cool atmosphere comes to flow out before heated up to 
coronal temperature ($T\sim 10^6$ K), as the gravity confinement weakens.

Many pioneering works have tried to reveal nature of red giant winds (e.g. 
Hartmann \& MacGregor 1980; Holzer, Fl\aa, \& Leer 1983; Charbboneau \& 
MacGregor 1995).  
However, the consecutive processes from the surface to the wind explained 
above have not been studied self-consistently;   
previous model calculations 
started from locations far above the photosphere in order to avoid 
difficulties arising from the huge density contrast. 
Moreover, they assume mechanical or phenomenological ``wave'' energies 
with somewhat ad hoc prescription of the dissipation.  

In this paper, we perform dynamical simulations of stellar winds from the 
photosphere to $\simeq 25\; R_{\star}$ in open magnetic field regions, 
where $R_{\star}$ is stellar radius. 
We give perturbations at the photosphere originating from the surface 
convection. 
These fluctuations excite waves that propagate upwardly. 
Among various modes the \Alfven wave travels to a sufficiently large 
distance to contribute to the acceleration and heating of the stellar wind. 
We simulate the propagation and dissipation of the waves by solving nonlinear 
MHD equations, and treat the heating and acceleration of the gas in a fully 
self-consistent manner without conventional heating functions 
(\S\ref{sec:dwv}). 
Wind properties, such as $\dot{M}$ and $v_{\infty}$, are determined from 
the surface conditions without ad hoc parameters. 
Another advantage of dynamical simulations 
is that we do not have to care about the critical (sonic) point(s), which is 
often problematic when constructing a transonic wind solution under 
the steady-state condition. A stable wind structure is automatically 
selected whether it is transonic or subsonic.

We already carried out simulations in the solar corona and wind 
\citep[][hereafter SI05; SI06]{si05,si06}. 
We showed that the coronal heating and solar wind acceleration in the open 
field regions are natural consequences of the 
photospheric perturbations; the \Alfven waves generated from the surface 
effectively dissipate in the corona by nonlinear mechanism to sufficiently 
heat and accelerate the solar wind in spite of the energy loss via 
radiative cooling, thermal conduction, and wave reflection. 
The aim of this paper is to study the evolution of the stellar winds from 
MS to RGB stages by extending this work to cool RGB stars. 
We investigate the stellar winds of a 1$M_{\odot}$ star from  
MS ({\it i.e.} present sun) to RGB across the DL, where $M_{\odot}$ is the 
solar mass. 
For comparison we also study a more massive RGB star with $3M_{\odot}$ 
both blueward and redward of the DL.  

\S\ref{sec:mod} presents our simulation method. In \S\ref{sec:esw} 
we show the results of the simulations, focusing on the evolution and 
time-dependent nature of the stellar winds. 
We discuss comparisons with related works as well as the uncertainties and 
limitations of our simulations in \S\ref{sec:dis}.   

\section{Simulation Set-up}
\label{sec:mod}
\subsection{Simulated Stars}

\begin{figure}
\figurenum{1} 
\epsscale{1.18}
\begin{center}
\plotone{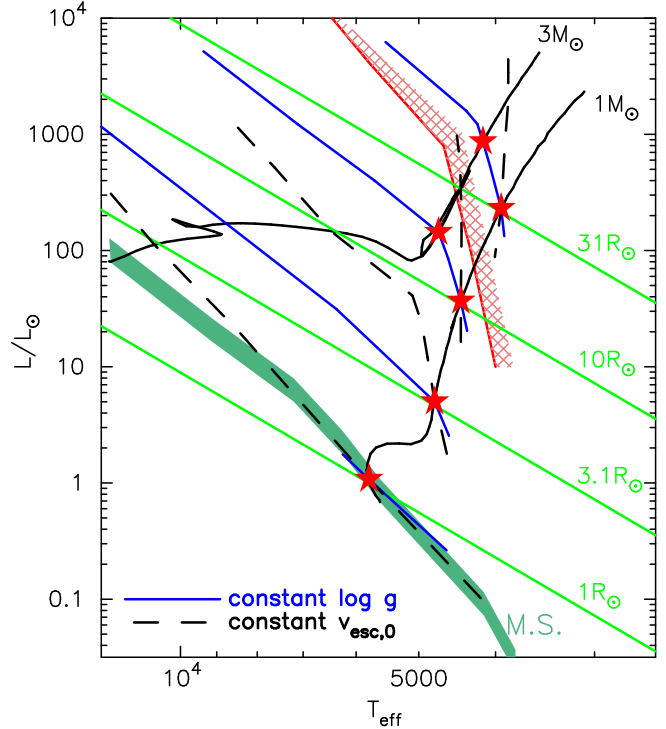}
\end{center}
\caption{Simulated stars in HR-diagram. The six red stars correspond to 
Models tabulated in Table \ref{tab:sttp}. The black solid lines are the 
evolutionary paths of 1 $M_{\odot}$ and $3\; M_{\odot}$ stars with the 
solar abundance \citep{gir00}. 
The red line with the cross hatch is the DL \citep{lh79}. 
The green solid lines 
indicate constant stellar radii, $R_{\star} = 1\; R_{\odot}$, $3.1\; 
R_{\odot}$, $10\; R_{\odot}$, and $31\; R_{\odot}$ from below. 
The blue lines indicate constant gravity, $\log g = 4.4$, 3.4, 2.4, 1.4 
from below. The black dashed lines indicate constant escape speeds 
from the stellar surfaces, $v_{\rm esc,0}= 617$, 347, 195, 110 km s$^{-1}$ 
from below. The dark green shaded region denotes the location of main 
sequence stars.}
\label{fig:hr}
\end{figure}

\begin{table*}
\begin{tabular}{|c|c|c|c|c|c|c|c|c|c|c|c|}
\hline
Model & $M_{\star}(M_{\odot})$ & $R_{\star}(R_{\odot})$ & $T_{\rm eff}$(K) 
& $\log g$ & $v_{\rm esc,0}$(km s$^{-1}$) & $\rho_{0}$(g cm$^{-3}$) 
& $\langle \delta v_0 \rangle$(km s$^{-1}$) & $\tau_{\rm min}$(min) 
& $\tau_{\rm max}$(min) & $R_1(R_{\star})$ & $R_2(R_{\star})$ \\
\hline
\hline
I & $1$ & $1$ & $5800$ & $4.4$ & 617 & $1.0\times 10^{-7}$ & 1.0 
& 1 & 32 & $1.01$ & $1.2$\\
\hline
II & $1$ & $3.1$ & $4800$ & $3.4$ & 347 & $5\times 10^{-8}$ 
& 1.5 & 9 & 290 & $1.031$ & $1.4$ \\
\hline
III & $1$ & $10$ & $4500$ & $2.4$ & 195 & 
$1.5\times 10^{-8}$ & 2.8 & 87 & 2800 
& 1.1 & 1.5 \\
\hline
IV & $1$ & $31$ & $3900$ & $1.4$ & 110 & 
$4.5\times 10^{-9}$ & 5.3 & 800 & 25600  
& $1.31$ & $1.8$\\
\hline
V & $3$ & $17.3$ & $4700$ & $2.4$ & 255 & 
$1.5\times 10^{-8}$ & 3.7 & 90 & 2900  
& $1.1$ & $1.5$ \\
\hline
VI & $3$ & $54.8$ & $4200$ & $1.4$ & 144 & 
$4\times 10^{-9}$ & 8.0 & 930 & 29900  
& $1.31$ & $1.8$ \\
\hline
\end{tabular}
\caption{Model properties. Tabulated from left to right are respectively 
stellar mass, $M_{\star}$, in $M_{\odot}$, stellar radius, $R_{\star}$, in 
$R_{\odot}$, effective 
temperature, $T_{\rm eff}$ (K), logarithm of surface gravity, $g=G M_{\star}/
R_{\star}^2$, escape speed at the surface, $v_{\rm esc,0}\equiv 
v_{\rm esc}(R_{\star}) 
=\sqrt{2G M_{\star}/R_{\star}}$, photospheric density, 
$\rho_{0}$(g cm$^{-3}$), root-mean-squared amplitude of transverse 
fluctuation at the photosphere, $\langle \delta v_0\rangle$ (km s$^{-1}$), 
minimum wave period, $\tau_{\rm min}$, maximum wave period, 
$\tau_{\rm max}$, and locations of flux tube expansions, $R_1$ and $R_2$.  }
\label{tab:sttp}
\end{table*}

We simulate the \Alfven wave-driven stellar winds of the various stars  
summarized in Table \ref{tab:sttp} and in 
Figure \ref{fig:hr}. 
First, we consider the $1\; M_{\odot}$ stars with different stellar radii, 
$R_{\star} = 1\; R_{\odot}$ ({\it i.e.} the present sun), $3.1\; R_{\odot}$, 
$10\; R_{\odot}$, and $31\; R_{\odot}$, where $R_{\odot}$ is solar radius, 
to study the evolution of the winds;   
note that the DL is between the latter two cases. 
We also simulate the stars with $3M_{\odot}$ near the DL to compare 
with the $1\; M_{\odot}$ cases.   
Here we adopt the stellar evolutionary tracks from \citet{gir00} to
determine the basic stellar parameters, such as $R_{\star}$ and effective 
temperatures, $T_{\rm eff}$, 
which are used to estimate the densities, $\rho_{0}$, at the photospheres 
and the photospheric amplitudes, $ \delta v_0$, of the surface convection 
(\S\ref{sec:phfl}). 

\begin{figure}
\figurenum{2} 
\epsscale{0.5}
\begin{center}
\plotone{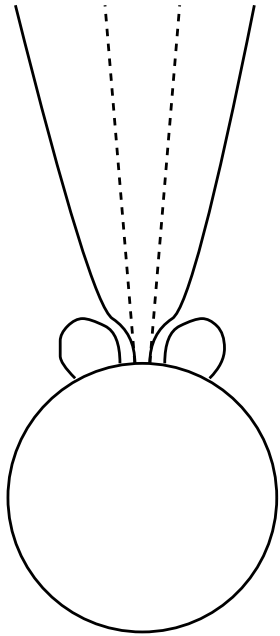}
\end{center}
\caption{Schematic picture of a magnetic flux tube that is super-radially 
open. 
The solid lines denote magnetic field lines. The open field region expands 
more rapidly than the spherical expansion shown by the dashed lines. 
Our simulations treat such super-radially open flux tubes.  
}
\label{fig:flxtb}
\end{figure}

\subsection{Simulation Details}
\label{sec:siml} 
We consider 1D flux tubes that are super-radially open. As shown in Figure 
\ref{fig:flxtb}, a certain fraction of the stellar surface is covered by 
closed loops so that open flux tubes diverge more rapidly than 
the radial expansion \citep{kh76}.

We inject velocity and magnetic field fluctuations originating from 
the surface convection at the photospheres, which are the inner boundaries 
of the simulations (\S\ref{sec:phfl}).
These fluctuations excite various waves, a fraction of which propagate 
upwardly. We treat both energy and momentum transfers in stellar winds, namely 
the propagation and dissipation of the waves and the consequent acceleration 
and heating 
of the surrounding gas, in a self-consistent manner.  We set the outer 
boundaries at $\simeq 25\; R_{\star}$ at which the outgoing condition of 
both materials and MHD waves is properly prescribed (SI06). 
In this paper we neglect centrifugal force, 
which is a reasonable assumption since 
low mass MS stars and RGB stars generally rotate slowly,  
and radiation pressure, 
which is also appropriate, where it might become 
important in more evolved asymptotic giant branch stars with dusts in 
the atmospheres.  

Radial field strength, $B_r$, 
is given by the conservation of magnetic flux as 
\begin{equation}
B_r r^2 f(r) = {\rm const.} ,
\end{equation}
where $f(r)$ is a super-radial expansion factor. 
A 1D flux tube is fixed by the field strength at the photosphere, 
$B_{r,0}$, and $f(r)$. 
We set $B_{r,0}$ and $f(r)$ based on our previous solar studies (SI05; SI06). 
While the average field strength is $1-10$ G on the solar photosphere, 
magnetic field lines 
are swept into inter-granulation lanes so that the localized strength 
becomes $\gtrsim 100$ G to a few kG. These field lines rapidly 
expand and becomes $\lesssim 10$G in the coronal region.  
Finally in the outer region ($\gtrsim$ a few $R_{\star}$) the field lines  
distribute almost radially. 

We use the the same $B_{r,0}=240$ G and the same total expansion 
factor, $f_{\rm tot}=240$, for all the Models. Note that these values give  
$B_{r,0}/f_{\rm tot}=1$ G, which is the total magnetic flux of the open 
field regions divided by the stellar surface. The average field strength 
becomes larger by adding the contributions from closed field regions 
(\S\ref{sec:unc}). 
We use the same functional form of the expansion as in SI06 
(see also Kopp \& Holzer 1976):
\begin{equation}
f(r) = \frac{f_{\rm 1,tot}\exp(\frac{r-R_1}{\sigma_1})+f_1}
{\exp(\frac{r-R_1}{\sigma_1})+1} 
\frac{f_{\rm 2,tot}\exp(\frac{r-R_2}{\sigma_2})+f_2}
{\exp(\frac{r-R_2}{\sigma_2})+1}
\end{equation}
where $f_1=1-(f_{\rm 1,tot}-1)\exp(\frac{R_{\star}-R_1}{\sigma_1})$ 
and $f_2=1-(f_{\rm 2,tot}-1)\exp(\frac{R_{\star}-R_2}{\sigma_2})$. 
A flux tube initially expands by a factor of $f_{\rm 1,tot}$ around 
$R_1 - \sigma_1 \sim R_1 + \sigma_1$, corresponding to the 'funnel' structure 
\citep{tu05} by closed loops, and followed by $f_{\rm 2,tot}$ times expansion  
around $R_2 -\sigma_2 \sim R_1 + \sigma_2$ due to the large scale (dipole) 
magnetic fields.  
The total super-radial expansion is defined as $f_{\rm tot} 
=f_{\rm 1,tot} f_{\rm 2,tot}$. We adopt $f_{\rm 1,tot} = 40$ and 
$f_{\rm 2,tot}=6$ in all the Models.
$R_1$ and $R_2$ of each Model are summarized in Table \ref{tab:sttp}. 
$R_1-R_{\star}$ is set to be (roughly) proportional to the scale height 
($\propto g^{-1}=(GM_{\star}/R_{\star}^2)^{-1}$; see Equation \ref{eq:isdns} 
for a more specific form) because closed loops are supposed to be taller 
in lower gravity stars if the surface 
magnetic field conditions are the same (but see \S\ref{sec:unc}).  
Accordingly $R_2$ is also set to be larger in more evolved stars. 
We adopt the 'widths' of the expansions, $\sigma_1 \approx 0.6
(R_1 -R_{\star}) $ and $\sigma_2 \approx 0.6 (R_2 -R_{\star}) $. 
Actually, the most important parameter that determines the stellar wind 
structure is $f_{\rm tot}$ and the other parameters, $R_1$, $R_2$, $\sigma_1$, 
$\sigma_2$, $f_{\rm 1,tot}$, and $f_{\rm 2,tot}$, only give small 
corrections.  

We dynamically solve ideal MHD equations with radiative cooling and thermal 
conduction. Conservation of mass, momentum (both radial and tangential 
components), and energy, and evolution of magnetic field are respectively 
expressed as  
\begin{equation}
\label{eq:mass}
\frac{d\rho}{dt} + \frac{\rho}{r^2 f}\frac{\partial}{\partial r}
(r^2 f v_r ) = 0 , 
\end{equation}
$$
\rho \frac{d v_r}{dt} = -\frac{\partial p}{\partial r}  
- \frac{1}{8\pi r^2 f}\frac{\partial}{\partial r}  (r^2 f B_{\perp}^2)
$$
\begin{equation}
\label{eq:mom}
+ \frac{\rho v_{\perp}^2}{2r^2 f}\frac{\partial }{\partial r} (r^2 f)
-\rho \frac{G M_{\star}}{r^2}  , 
\end{equation}
\begin{equation}
\label{eq:moc1}
\rho \frac{d}{dt}(r\sqrt{f} v_{\perp}) = \frac{B_r}{4 \pi} \frac{\partial} 
{\partial r} (r \sqrt{f} B_{\perp}).
\end{equation}
$$
\rho \frac{d}{dt}\left(e + \frac{v^2}{2} + \frac{B^2}{8\pi\rho} 
- \frac{G M_{\star}}{r}\right) 
+ \frac{1}{r^2 f} 
\frac{\partial}{\partial r}\left[r^2 f \left\{ (p + \frac{B^2}{8\pi}) v_r  
\right. \right.
$$
\begin{equation}
\label{eq:eng}
\left. \left.
- \frac{B_r}{4\pi} (\mbf{B \cdot v})\right\}\right]
+  \frac{1}{r^2 f}\frac{\partial}{\partial r}(r^2 f F_{\rm c}) 
+ q_{\rm R} = 0,
\end{equation}
\begin{equation}
\label{eq:ct}
\frac{\partial B_{\perp}}{\partial t} = \frac{1}{r \sqrt{f}}
\frac{\partial}{\partial r} [r \sqrt{f} (v_{\perp} B_r - v_r B_{\perp})], 
\end{equation}
where $\rho$, $\mbf{v}$, $p$, $\mbf{B}$ are density, velocity, pressure, 
and magnetic field strength, respectively, and subscript 
$r$ and $\perp$ denote radial and tangential components; 
$\frac{d}{dt}$ and $\frac{\partial}{\partial t}$ denote Lagrangian and 
Eulerian derivatives, respectively; 
$e=\frac{1}{\gamma -1}\frac{p}{\rho}$ is specific energy and we assume 
the equation of state for ideal gas with a ratio of specific heat, 
$\gamma=5/3$; 
$G$ is the 
gravitational constant; 
$F_{\rm c} = \kappa_0 T^{5/2} 
\frac{dT}{dr}$ is thermal conductive flux by Coulomb collisions, where 
$\kappa_0 \approx 10^{-6}$ in c.g.s unit;  
$q_{\rm R}$ is radiative cooling. 
In this paper we assume the solar elemental abundance for the radiative 
cooling term. 
In sufficiently low density regions with $\rho \le 5\times 10^{-17}$ g 
cm$^{-3}$ we use optically thin radiative loss, $q_{\rm R}=
n_p n_e \Lambda$, where $n_p$ and $n_e$ are proton and electron number 
densities and $\Lambda$ is the tabulated cooling function by \citet{LM90}. 
In denser regions, which generally corresponds to low chromosphere, 
we adopt empirical radiative 
cooling, $q_{\rm R} = 4.5\times 10^9 \rho$ (erg cm$^{-3}$s$^{-1}$) 
derived from observations of the solar chromosphere 
\citep{aa89,mor04} to take into account the optically thick effect. 
For simplicity's sake we switch off the cooling if temperature drops 
below 3000 K, which sometimes happens in the RGB stars. 
We might have to take into account effects of neutral atoms or dusts in such 
low temperature, though we do not do explicitly in this paper 
(see \S\ref{sec:lim}).

We adopt the second-order MHD-Godunov-MOCCT scheme (Sano \& Inutsuka 2007 
in preparation; see also SI06) to update the physical quantities. 
In this scheme each cell boundary is treated as a discontinuity, and for the 
time evolution we solve the nonlinear Riemann shock tube problem with 
magnetic pressure by using the Rankine-Hugoniot relations. 
Therefore, entropy generation, 
namely heating, is automatically calculated from the shock jump condition. 
A great advantage of our code is that no artificial viscosity is required 
even for strong MHD shocks; numerical diffusion is 
suppressed to the minimum level for adopted numerical resolution.

\subsection{Surface Conditions}
\label{sec:phfl}
We describe the conditions of the stellar photospheres, which are the inner 
boundaries of the simulations. 
\subsubsection{Density}
The density, $\rho_0$, at photosphere (defined as optical depth $\approx 1$) 
is subject to stellar parameters through various opacity effects in the 
atmosphere. 
The gas pressure, $p_0$, at photosphere has a positive dependence on gravity, 
$p_0 \propto g^{0.6}$, for a given $T_{\rm eff}$ 
(\S9 of Gray 1992).  
$p_0$ has a negative dependence on $T_{\rm eff}$, and in the cool star 
condition (4000 K$\lesssim T_{\rm eff} \lesssim 6000$ K) 
the relation is roughly 
$p_0\propto T_{\rm eff}^{-2}$ (Figure 9.21 of Gray 1992).   
Then, $\rho_0$($\propto p_0/T_{\rm eff}$) can be scaled as 
\begin{equation}
\rho_0 \propto g^{0.6} T_{\rm eff}^{-3}. 
\label{eq:sfprs}
\end{equation}
We firstly set $\rho_0=10^{-7}$ g cm$^{-3}$ on the sun and $\rho_0$'s 
in the other cases are determined by this scaling relation; $\rho_0$'s in  
Table \ref{tab:sttp} are thus given\footnote{$\rho_0$'s in Table 
\ref{tab:sttp} are slightly different ($<10$\%) from the values derived from 
Equation (\ref{eq:sfprs}) This is because $\rho_0$'s are set by the model 
atmosphere that is used to derive the approximated scaling of 
Equation (\ref{eq:sfprs}).}.

\subsubsection{Wave Amplitude}
The fluctuation amplitude, $\delta v_0$, at the photospheres can be estimated 
from convective flux.  
The generation of sound waves on late-type stars have been extensively studied 
(Lighthill 1952; Stein 1967; de Loore 1970; Renzini et al. 
1977; Bohn 1984; Shibahashi 2005). Recently, numerical experiments of surface 
convection are also carried out \citep{ste04}.         
These works show a similar qualitative trend that the excited acoustic flux, 
$F_{\rm a,0}$(erg cm$^{-2}$s$^{-1}$), 
has a negative dependence on surface gravity and a positive dependence on 
stellar effective temperature. In this paper we adopt the relation derived by 
\citet{ren77},  
\begin{equation}
F_{\rm a,0} (\approx \rho_0 \langle \delta v_{0}^2\rangle c_s) 
\propto g^{-0.7} T_{\rm eff}^{12}, 
\label{eq:acfsc}
\end{equation} 
where $\delta v_{0}$ is wave amplitude, 
$c_s$ is sound speed, and $\langle \rangle$ denotes time-average.

Strictly speaking, $\delta v_0$ should be longitudinal (parallel to 
the field line; {\it i.e.} radial direction) component. 
In this paper we assume both transverse and longitudinal components have 
the same amplitude at the photosphere.  
This assumption is roughly satisfied at least on the sun because 
observation of solar granulations shows that both transverse and longitudinal 
amplitudes are similar, 
$\approx 1-2$ km s$^{-1}$ \citep{hgr78}. 
We set $\langle \delta v_0\rangle = 1.0$ km s$^{-1}$ in the present sun 
(Model I), where $\langle \delta v_0\rangle \equiv
\sqrt{\langle \delta v_{\parallel,0}^2 \rangle}$. 
For the other stars, we can use the following scaling, 
\begin{equation}
\langle \delta v_0\rangle \propto g^{-0.65} T_{\rm eff}^{7.25}, 
\label{eq:dvscl}
\end{equation}
which is derived from Equations (\ref{eq:sfprs}) and (\ref{eq:acfsc}) 
(note $c_s\propto T_{\rm eff}^{1/2}$). 
$\langle \delta v_0\rangle$'s in Table \ref{tab:sttp} are determined 
by this relation\footnote{$\langle \delta v_0\rangle$'s in Table 
\ref{tab:sttp} are also slightly different from those based on Equation 
(\ref{eq:dvscl}). This is again because we are using the model atmosphere
when setting $\delta v_0$.}. 

\subsubsection{Wave Spectrum}
The typical period, $\tau$, of generated waves is also related to stellar 
parameters. 
$\tau$ can be scaled by eddy (granulation cell) turn-over timescale, 
$\sim l/c_s$, where 
$l$ is a typical granulation size, if stochastic processes of the convection 
dominate in the wave generation. 
$l$'s are different in different stars, and are probably proportional to 
the pressure scale heights, $H_p \approx c_s^2/g$.  
Then, we have
\begin{equation}
\tau \sim l/c_s \propto H_p/c_s \sim c_s/g \propto g^{-1} T_{\rm eff}^{0.5}, 
\label{eq:wpsc}
\end{equation}
which is consistent with recent numerical simulations ({\it e.g.} 
Stein et al. 2004).

We use Equation (\ref{eq:wpsc})
to give the realistic spectra of fluctuations at the photosphere. 
In this paper, we adopt a broadband power spectrum of $P(\nu)\propto \nu^{-1}$ 
with respect to frequency, $\nu=1/\tau$, where the normalization is given by 
$\langle \delta v_0^2 \rangle =\int_{\nu_{\rm min}}^{\nu_{\rm max}} d\nu 
P(\nu)$.  For the present sun, we set $\tau_{\rm min} (= 1/\nu_{\rm max}) 
= 1$ min and $\tau_{\rm max} (= 1/\nu_{\rm min}) = 32$ min, giving 
the (logarithmic) mean value $\approx 5$ min, which dominates the solar 
oscillations. 
$\tau_{\rm min}$'s and  $\tau_{\rm max}$'s of the other stars can be 
determined by the scaling relation of Equation (\ref{eq:wpsc}) as shown 
in Table \ref{tab:sttp}.  
It should be noted that a shape of $P(\nu)$ affects wind structure 
little as long as the same $\tau_{\rm min}$ and  $\tau_{\rm max}$ are 
adopted (SI06). 

\subsubsection{Wave Mode}
In this paper we only consider the open field regions in which the 
(unperturbed) magnetic fields are oriented to the vertical direction at 
the surfaces. In such a situation, if we restrict 
waves traveling in the vertical direction, the transverse 
components of surface fluctuations produce \Alfven waves, while the 
longitudinal component excites compressive waves, e.g. acoustic waves, or 
more generally, magnetosonic waves. 
The uncompressive character enables the \Alfven waves to propagate long 
distance.  
On the other hand, the longitudinal waves 
are quite dissipative because 
both nonlinear shock formation \citep{suz04} and collisionless dissipation 
\citep{shl06} are effective.    
Thus, the compressive waves cannot travel a sufficiently 
long distance to contribute to the acceleration of the winds \citep{suz02}. 
Actually, we tested a role of 
longitudinal fluctuations in several models and found that its effect is 
limited in the regions near the surfaces  ($\ll 2\; R_{\star}$). 
Therefore, we only show the results of transverse fluctuations at the 
photospheres in this paper. 

\section{Results}
\label{sec:esw}
\begin{figure}
\figurenum{3} 
\epsscale{1.}
\begin{center}
\plotone{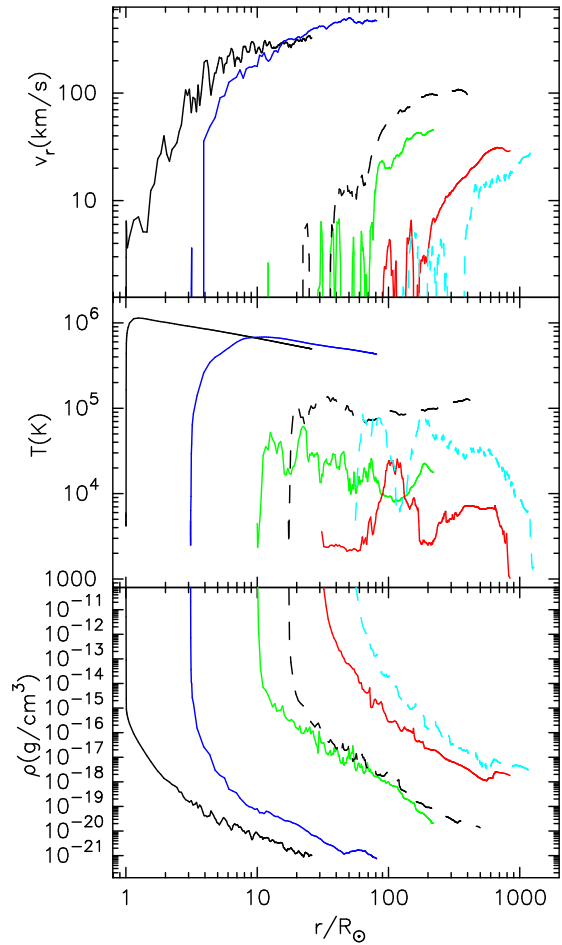}
\end{center}
\caption{Time-averaged stellar wind structure of the six Models. 
From the top to the 
bottom, temperature, $T$ (K), radial outflow velocity, $v_r$ (km s$^{-1}$),  
and density, $\rho$(g cm$^{-3}$), are plotted. The solid lines are the results 
of the 1 $M_{\odot}$ stars; the black, blue, green, and red lines are the 
results of Models I, II, III, and IV, respectively, 
The dashed lines are the results of the $3\; M_{\odot}$ stars; the black and 
light-blue lines are the results of Models V and VI.  
}
\label{fig:wdst}
\end{figure}
We initially set up the static and cool atmospheres with the 
photospheric temperatures; {\it neither stellar winds nor hot coronae are 
set up in advance}. 
We start the simulations by injecting the transverse fluctuations from the 
photospheres, which excite \Alfven waves 
(see the {\it electronic edition} for the mpeg file of Model VI which is 
the movie version of Figure \ref{fig:mpw}).
The atmospheres are accelerated as transonic winds in all the six
Models by the dissipation of the \Alfven waves (\S\ref{sec:dwv}). 
This shows that the transonic outflow is 
a stable and natural consequence of the \Alfven wave-driven winds 
of both MS and RGB stars.     
The stellar winds are settled down to the quasi-steady states after $\gtrsim$ 
twice of the sound crossing times of the entire simulation boxes 
(the sound speeds are slower than the \Alfven speeds in our simulations), 
although the RGB cases (Models III-VI) show large fluctuations in the winds 
by thermal instability (\S\ref{sec:tdp}). 

\begin{figure}
\figurenum{4} 
\epsscale{1.}
\begin{center}
\plotone{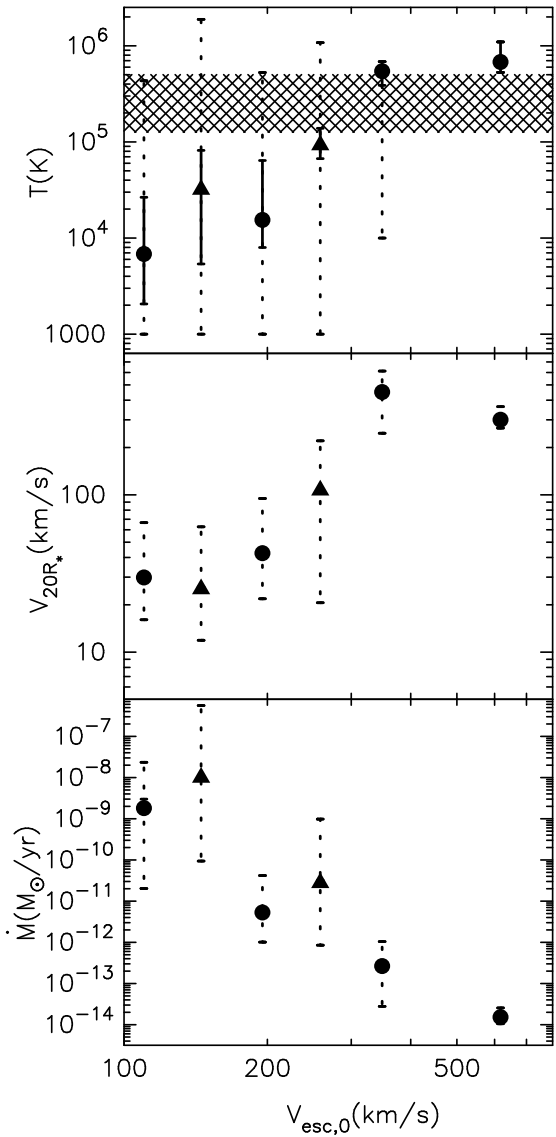}
\end{center}
\caption{Typical wind parameters of the six Models as functions of 
surface escape velocity, $v_{\rm esc,0}$ (km s$^{-1}$). The circles and 
triangles represent the time-averaged values of the $1\; M_{\odot}$ 
(Models I - IV) and 
$3\; M_{\odot}$ (Models V \& VI) stars, respectively. 
From the top to the bottom temperature, $T$ (K), outflow velocity, 
evaluated at $r=20\; R_{\star}$ 
are plotted. 
$T$ (K) is averaged over the range of $1.5\; R_{\star}\le r \le 
15\; R_{\star}$. 
The solid error bars in the top panel correspond to the maximum and minimum 
values of the time-averaged structure in this range. 
The dashed error bars are the results of the maximum and 
minimum values derived from the snap-shot results (\S \ref{sec:tdp}). 
The hatched region in the top panel indicates the thermally unstable region. 
}
\label{fig:chp}
\end{figure}
  
First we would like to briefly mention the main results of our simulations. 
Figure \ref{fig:wdst} presents the structure of temperature, $T$ (K) 
(top), outflow velocity, $v_r$ (km s$^{-1}$) (middle), and density, 
$\rho$ (g cm$^{-3}$) (bottom) 
on radial distances from the stellar centers in unit of 
$R_{\odot}$.  The quantities are averaged over  
sufficiently longer durations than $\tau_{\rm max}$ of each Model. 
Figure \ref{fig:chp} shows the typical wind parameters, temperature, $T$(K) 
in the atmosphere (top), 
wind speed, $V_{20R_{\star}}$ (km s$^{-1}$), at 20 $R_{\star}$ (middle), 
and mass loss rate, $\dot{M}=4\pi (\rho r^2 v)_{20R_\star}$
($M_{\odot}$ yr$^{-1}$), evaluated 
at 20 $R_{\star}$ (bottom), as functions of $v_{\rm esc,0}$; evolved stars
are in the left side. 
Note that the super-radial expansions of the flux tubes finish in $r<
3R_{\star}$ so the flux tubes are radially oriented at $r=20R_{\star}$.  
In the top panel, we are averaging $T(r)$ in 1.5 - 15 $R_\star$. 
The solid error bars correspond to the temperature ranges of the time-averaged 
structure in this region. 
The dashed error bars in all the panels correspond to the maximum and minimum 
values from the snap-shot results (\S\ref{sec:tdp}). 

One of the most important results is the disappearance of {\it steady} hot 
coronae in the RGB stars (Models III - VI). It is clearly seen in the top 
panel of Figure \ref{fig:chp} that the average temperature drops suddenly 
from $T\simeq 7\times 10^5$K in the sub-giant star (Model II) 
to $T\le 10^5$K in the RGB stars.  The large dashed error bars in the same 
panel indicate that the temperatures of the RGB atmospheres show large 
variations. One may also see in the middle and bottom panels that the 
velocities and mass loss rates of the RGB stars largely fluctuate. 
Particularly, $\dot{M}$'s of the RGB stars vary 2-4 orders 
of magnitude. Therefore, the red giant winds are highly structured, in 
contrast to the steady coronal winds from the MS and sub-giant stars.     

The middle panel of Figure \ref{fig:wdst} shows that nearly-static 
regions are formed above the photospheres in the RGB stars, and 
the acceleration of the winds essentially starts from several stellar radii. 
Accordingly, their wind speeds in the outer regions are considerably slower 
than the escape velocities at the surfaces (the middle panel of Figure 
\ref{fig:chp}), which is consistent with the observed trend ({\it e.g.} 
Dupree). 
The mass loss rate rapidly increases from MS to RGB. The average $\dot{M}$'s 
of the stars with $\log g=1.4$ are $10^{-9}-10^{-8}\; M_{\odot}$yr$^{-1}$, 
which are comparable to the observed level \citep{js91}.

\subsection{Disappearance of Steady Coronae}
\label{sec:dsp}
We explain the mechanism of the disappearance of the steady corona 
in more detail. 
The main reason is that the sound speed ($\approx 150$ km s$^{-1}$) 
of $\approx 10^6$ K plasma exceeds the escape speed, 
$v _{\rm esc}(r)=\sqrt{2G M_{\star}/r}$, at $r \gtrsim$ a few $R_{\star}$; 
the hot corona cannot be confined by the gravity any more in the atmospheres 
of the RGB stars.  
Therefore, the material flows out before heated up to coronal temperature.  
However, 
the temperature gap at $v_{\rm esc,0}\simeq 300$ km s$^{-1}$ (the 
top panel of Figure \ref{fig:chp}) cannot be explained only by this reason. 


Thermal instability plays an important role in the drop of the temperature. 
The radiative cooling function of the optically thin gas with the solar 
abundance \citep{LM90} shows a decreasing trend on temperature in $T>10^5$ K, 
indicating that gas is less cooled as it is heated up. 
Then, thermal instability sets in and gas in this temperature range cannot 
stably exist \citep{ki02}. In solar and stellar coronal 
situations, however, thermal conduction plays a role in stabilization; drastic 
heating to $T\sim 10^7$ K is suppressed by the downward conduction from 
the upper corona to the lower chromosphere \citep{ham82}.   
Then, gas with $T\gtrsim 10^6$ K becomes thermally stable, while 
gas with $10^5\; {\rm K} < T \lesssim 10^6$ K, which is shown by 
the hatched region in Figure \ref{fig:chp}, remains unstable.

Let us think about comparison of  $c_{\rm s}$ and $v_{\rm esc}$ at a few 
$R_{\star}$ of the $1\; M_{\odot}$ stars. 
In Models I and II, $v_{\rm esc}$ exceeds $c_{\rm s}$ ($\simeq 
150$ km s$^{-1}$) of $10^6$ K plasma. 
Under such a condition the coronal temperatures are determined by energy 
balance among heating, radiative cooling, and 
downward thermal conduction in the transition region which divides 
the chromosphere and the corona \citep{ham82}. 
However, in Model III($\log g=2.4$) $v_{\rm esc}(3R_{\star}) 
\sim 110$ km s$^{-1}$, and in Model IV($\log g=1.4$) 
$v_{\rm esc}(3R_{\star}) \sim 60$ km s$^{-1}$, 
which correspond to sound speeds of plasmas with $T \simeq 6\times 10^5$ K 
and $\simeq 2\times 10^5$ K, 
respectively. Therefore, the gas streams out before heated up to 
the stable temperature ($\gtrsim 10^6$ K) because of the insufficient 
confinement by the gravity. Those temperatures are thermally unstable, 
and then, the actual temperatures become lower than the gravity 
limited values.  
This is the reason why the average temperatures of Models III --- VI are not 
$10^5-10^6$ K but $< 10^5$ K.  
In conclusion, the thermal instability, in addition to the gravity effect, 
yields the sharp decline of the temperature from $T \gtrsim 10^6$ K 
to $T\lesssim 10^5$ K with the stellar evolution. 

The results (Figures \ref{fig:wdst} \& \ref{fig:chp}) show that the 
temperatures in the winds of the $3\; M_{\odot}$ RGB stars are 
systematically higher than those of the $1\; M_{\odot}$ counterparts.
This is because the energy inputs from the surfaces are larger in 
more massive stars. 
When fixing surface gravity, a star with a larger mass has higher effective 
temperature, which gives a larger photospheric amplitude 
(Equation \ref{eq:dvscl}). Then, more energy is injected from the surface 
so that the larger heating is achieved. 

\subsection{Structured Red Giant Winds}
\label{sec:tdp}
After the disappearance of the steady corona, the atmospheres and 
winds of the RGB stars show time-dependent behaviors because of the thermal 
instability. In this section, we examine the snap-shot structure of the 
red giant winds. 
We firstly show the results of Model VI ($3\; M_{\odot}$), which is 
redward of the DL but more active than the $1\; M_{\odot}$ counterpart 
(Model IV). After that, we discuss the other RGB cases.


\begin{figure}
\figurenum{5} 
\epsscale{1.}
\begin{center}
\plotone{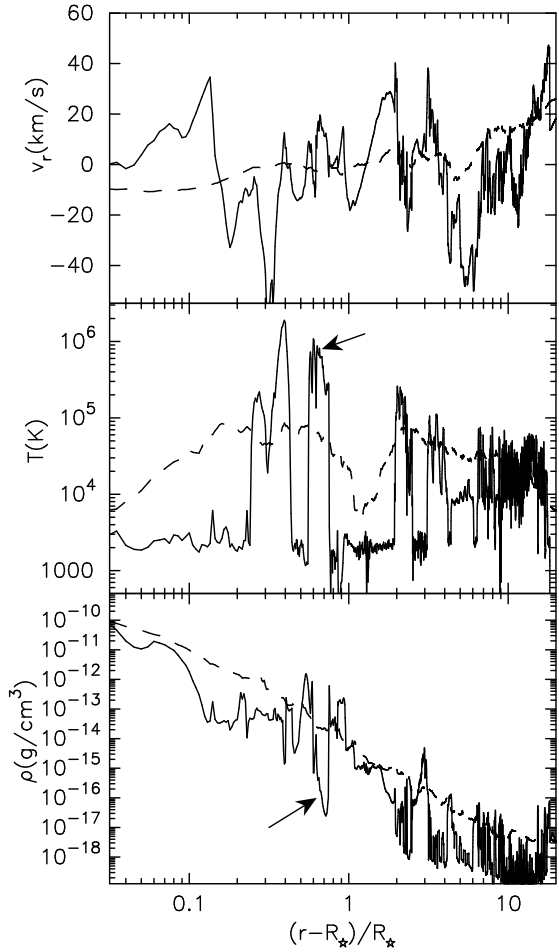}
\end{center}
\caption{Snap-shot wind structure (solid) of Model VI at 
$t=6909$(hr), compared 
with the time-averaged structure (dashed). 
From the top to the bottom,  $v_r$ (km s$^{-1}$), $T$ (K), and 
$\rho$ (g cm$^{-3}$) are plotted on $(r-R_{\star})/R_{\star}$. 
The arrows in the middle and bottom 
panels show the magnetized hot bubble which we inspect in \S\ref{sec:mhb} and 
Figure \ref{fig:mhb}. The mpeg movie which shows this simulation from 
$t=0$ to 9800 hr is available in the {\it electronic edition.} The movie 
file presents $\langle v_{\perp} \rangle$ (km s$^{-1}$) in addition to 
$v_r$ (km s$^{-1}$), $T$ (K), and $\rho$ (g cm$^{-3}$). 
}
\label{fig:mpw}
\end{figure} 

Figure \ref{fig:mpw} shows the snap-shot wind structure of Model VI 
at $t = 6909$(hr) (solid) in comparison with the time-averaged structure 
(dashed) that is the same as in Figure \ref{fig:wdst}. 
The mpeg movie of the simulation 
is also available in the  {\it electronic edition}.   
Figure \ref{fig:mpw} illustrates that the simple picture of layered 
atmosphere, photosphere -- chromosphere -- transition region -- corona -- 
wind from below, does not hold in the RGB star. 
A characteristic feature is that a number of hot bubbles with low densities 
are distributed in cool background material. 
For example, in an inner region ($r-R_{\star} < R_{\star}$) 
the two bubbles with the peak temperatures $>10^6$ K are seen; 
in $R_{\star} < r-R_{\star} 
\lesssim 10\; R_{\star}$ a couple of warm bubbles with $T\gtrsim 10^5$ K are 
formed. The hot and warm bubbles and the cool background materials 
are connected by the transition regions, at which the temperature 
drastically changes because of the thermal instability.

The densities of the hot bubbles are lower than the ambient media to 
satisfy the pressure balance. As seen in the bottom panel of Figure 
\ref{fig:mpw} the density fluctuates typically by 2-3 orders of magnitude 
in the wind, which is related to the fact that the gas mainly consists 
of hot ($>10^6$ K) and cool ($<10^4$ K) components. 
Strictly speaking, however, magnetic pressure also needs to be taken into 
account for the force balance (\S\ref{sec:mhb}).  
The outflow speed fluctuates as well
to fulfill the mass conservation relation. 
The red giant wind is not a steady outward stream but an outflow 
consisting of many small-scale structures. 
Our simulation shows that both hot plasma and cool chromospheric wind 
coexist in the RGB star, which is consistent with observations of hybrid 
stars \citep{hdr80,har95,ayr98}.


\subsubsection{Magnetized Hot Bubble}
\label{sec:mhb}

\begin{figure}
\figurenum{6} 
\epsscale{1.}
\begin{center}
\plotone{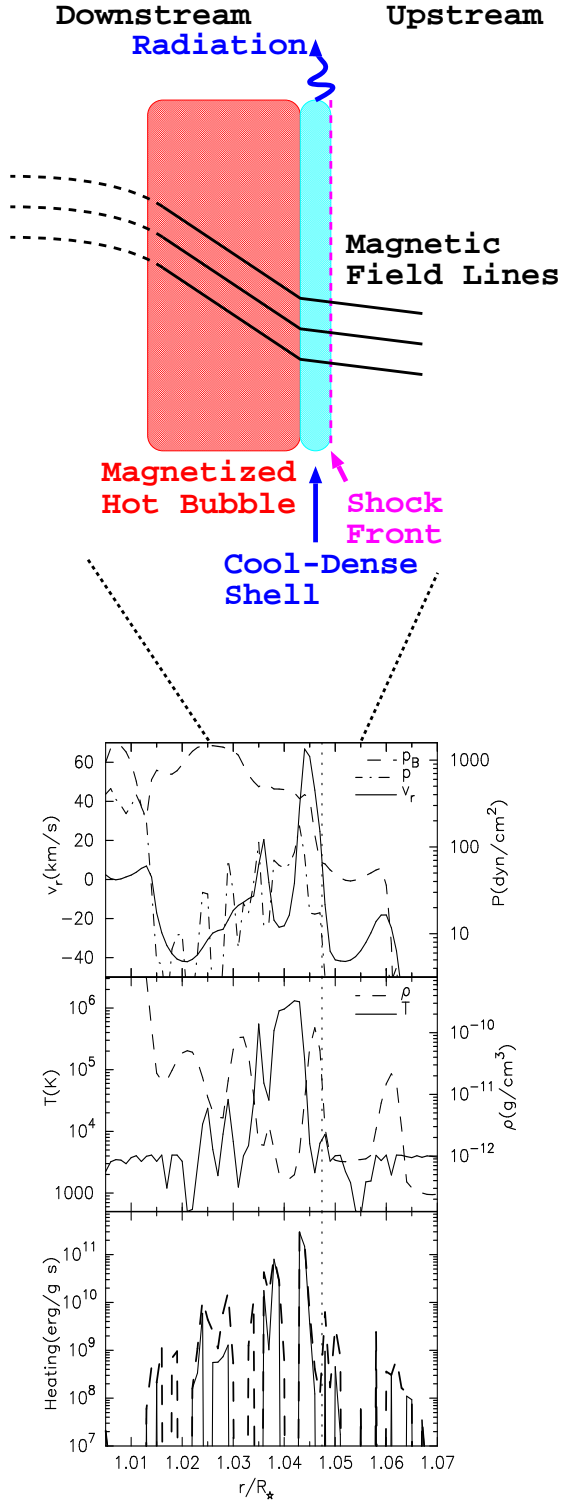}
\end{center}
\caption{Zoomed-in structure of the magnetized hot bubble 
at $t=6755$(hr) ({\it lower panels}), which is the initial phase of the 
bubble indicated by the arrows in Figure \ref{fig:mpw}, and schematic 
picture of a magnetized hot bubble ({\it upper cartoon}).  
The dotted lines in the lower panels show the location of the shock front. 
What are plotted in the lower three panels are as follows: 
{\it top} : The dashed, dot-dashed, and solid lines are 
magnetic pressure, $p_{\rm B}=B_{\perp}^2/8\pi$, gas pressure, $p$, 
and radial velocity, $v_r$.  
The values of $p$ and $p_{\rm B}$ are on the right axis 
(dyn cm$^{-2}$) and the value of $v_r$ is on the left axis (km s$^{-1}$). 
{\it middle} : The solid and dashed lines are temperature (K; left axis) and 
density (g cm$^{-3}$; right axis).  
{\it bottom} : The vertical axis shows heating rate (erg g$^{-1}$s$^{-1}$). 
The solid line is the 
net heating derived from the change of specific heat and the dashed line 
is the pure MHD heating which excludes the effect of radiative cooling 
and thermal conduction from the net heating.  
}
\label{fig:mhb}
\end{figure}

How are these hot bubbles formed ? How are the high temperatures 
maintained ?  
We study the formation and evolution of bubbles by inspecting  
one typical hot bubble that is indicated by the arrows in 
Figure \ref{fig:mpw}. This bubble is initially created at $r=
1.04\; R_{\star}$ at $t=6755$ hr. It comoves outwardly with the background 
flow and eventually cooled down to $\approx 10^4$ K at $t\simeq 7070$ hr 
at $r=1.6\; R_{\star}$ (see the mpeg movie of Figure \ref{fig:mpw}  
available in the {\it electronic edition}).

Figure \ref{fig:mhb} presents the zoomed-in structure just after 
the formation of the bubble (the lower panels) with 
schematic cartoon (top).  
The location of the shock front, which plays a role in the heating of 
the bubble, is indicated by the dotted line at 
$r=1.047\; R_{\star}$ in the panels. 
The top and middle panels show  
$p_{\rm B}$($=B_{\perp}^2/8\pi$; magnetic pressure), 
$p$, $v_r$, $T$ and $\rho$.  In the bottom panel, we present heating rate 
per mass. The solid line indicates the net heating rate that is calculated 
from local change of specific heat, $\partial e/\partial t$, and the 
dashed line shows the heating by the only MHD process which excludes 
radiative cooling and thermal conduction from $\partial e/\partial t$.


As clearly seen in the bottom panel the heating rate is not smoothly 
distributed but largely fluctuates because the heating mainly occurs at 
the steepened fronts of waves (see \S \ref{sec:dwv}). 
Radiative cooling is also important; 
in some portions the net heating becomes negative (the solid 
lines disappears) in spite of the positive MHD heating (dashed line), 
because the radiative cooling exceeds the MHD heating. 
This large variation of the heating rate by waves and cooling
triggers the multi-temperature stellar wind. 

$p_{\rm B}$ increases at the shock front from the upstream (right) side to 
the downstream (left) side, which indicates that 
this is a fast MHD shock.
One may find the density peak just behind the shock front. 
This is because an effective ratio 
of specific heats is small $\gamma_{\rm eff} \lesssim 1.1$ owing to the 
thermal conduction; the density jump at the shock becomes large 
in small $\gamma_{\rm eff}$ circumstances so that the material should be 
concentrated in the narrow region behind the shock to satisfy the mass 
conservation \citep{kr05}. 
As a result, the radiative cooling dominates the MHD heating owing to the 
high density, and this region is cooled down to form a cool dense shell
(the bottom panel of Figure \ref{fig:mhb}). 

Behind this cool dense shell, the hot region, which was heated up 
by the fast MHD shock, remains.  
The fast MHD shock plays an important role in maintaining this hot bubble. 
Because the magnetic field is generated by the shock, the magnetic pressure 
in the bubble is larger than in the ambient gas. Therefore, the hot bubble 
is not supported by the gas pressure but by the magnetic pressure. 
Even though the gas pressure decreases 
by an occasional cooling, the bubble can 
avoid being compressed. The density is kept low to reduce the radiative loss 
as well as to increase the heating rate per mass. 
The hot bubble 
survives much longer time than ambient (average) cooling time 
until the magnetic topology changes;     
the high temperature is maintained around 315hr, while a 
typical cooling time, $\tau_{\rm cool}$, of gas with $T=10^6$K ($\Lambda 
\approx10^{-22}$erg cm$^3$s$^{-1}$; see \S\ref{sec:siml}) is 
\begin{equation}
\tau_{\rm cool} =\frac{\rho e}{q_{\rm R}} \approx 1\; {\rm min}\left(
\frac{\rho}{10^{-13}{\rm g\; cm^{-3}}}\right)^{-1}
\end{equation} 
To summarize, hot bubbles are initially created by the variations of 
heating rate, and their high temperature is maintained thanks to the 
support by magnetic pressure.

An intriguing concept of magnetized hot bubbles is that hot plasma 
can be heated up even in open field regions. 
It has been generally considered that closed loops are necessary to 
confine hot plasma in the atmosphere of a RGB star since the gravity is not 
sufficiently strong. 
However, our simulations show that the hot bubble can exist 
rather long time ($\sim 10$ days) by magnetic topology, {\it i.e.} a fast MHD 
shock, in open field regions. 
It is supposed that such hot bubbles are ubiquitously created 
and they are possible sources of continuous soft X-ray 
emissions in hybrid stars as discussed later (\S\ref{sec:rdl}). 

\subsubsection{Time-Variation of Wind Parameters}

\begin{figure}
\figurenum{7} 
\epsscale{1}
\begin{center}
\plotone{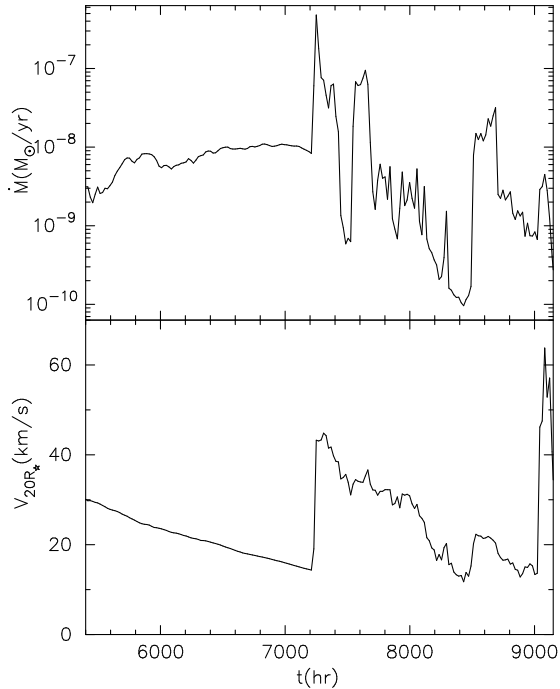}
\end{center}
\caption{Mass loss rate (upper) and radial velocity (lower) of the 
stellar wind of Model VI as functions of time. Both quantities are evaluated 
at $r=20\; R_{\star}$. 
}
\label{fig:mdtdp}
\end{figure} 

Because the red giant wind consists of many small-scale structures, 
the mass loss rate and the outflow speed in the outer region 
also show time-dependent behaviors. 
Figure \ref{fig:mdtdp} presents $\dot{M}$ ($M_{\odot}$ yr$^{-1}$) (upper) 
and $v_{20\; R_{\star}}$ (km s$^{-1}$) (lower). 
The upper panel shows $\dot{M}$ varies more than three orders of 
magnitude from $1\times 10^{-10}$ to $5\times 10^{-7}$ $M_{\odot}$ yr$^{-1}$. 
The lower panel also shows that $v_{20\; R_{\star}}$ largely 
varies from $10$ km s$^{-1}$ to $60$ km s$^{-1}$. 

The jumps are seen in $\dot{M}$ and $v_{20R_{\star}}$ at $t=7200, 8750$, 
and 9020 hr. 
These jumps indicate that dense blobs pass at $r=20\; R_{\star}$ at those 
times.
We examine one of the blobs which passes at $t=7200$ hr (This blob is seen in 
Figure \ref{fig:mpw} as well). 
This is a shock wave originally formed in an inner region. 
The shock amplitude has been initially small, but it grows as the shock moves 
outward by sweeping-up a number of smaller dense blobs as well as by 
amplification in the stratified atmosphere with the decreasing density. 
Finally the compression ratio of the shock increases to more than 10.  
(This value exceeds the adiabatic limit $=4$ since our simulations, 
which take into account cooling and thermal conduction, are not adiabatic.)
Before this shock arrives, $\dot{M}$ is almost constant and $v_{20R_{\star}}$ 
shows smoother feature. After the shock reaches $20R_{\star}$, 
both $\dot{M}$ and $v_{20R_{\star}}$ show fluctuating behaviors. 
This is because the downstream region contains larger perturbation 
than the upstream region.

\subsubsection{Radiative Flux}
\label{sec:rdl}
\begin{figure}
\figurenum{8} 
\epsscale{1.}
\begin{center}
\plotone{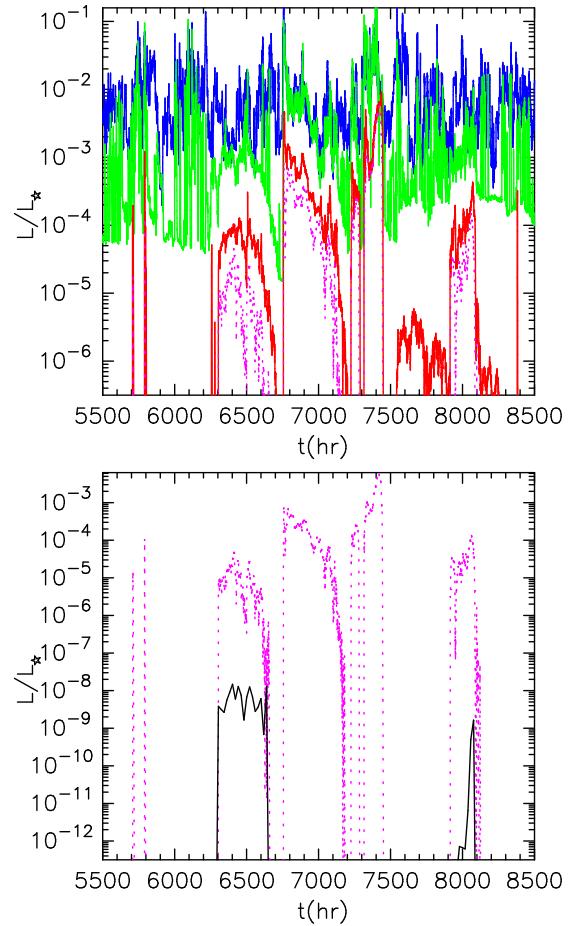}
\end{center}
\caption{{\it Upper}: 
Normalized radiative fluxes, $L/L_{\star}$, of cool ($T\le 2\times 
10^4$ K; blue), warm ($2\times 10^4 < T < 5\times 10^5$ K; green), and hot 
($T\ge 5\times 10^5$ K; red) components of Model VI as functions of time 
elapsed from the beginning of the simulation.  The radiative flux of 
$T\ge 10^6$ K gas, which is a part of the hot component, is also plotted 
(dashed line). 
{\it Lower}: Radiative flux of the soft X-ray ($T\ge 10^6$ K) 
that takes into account the 
absorption by the cool wind materials (black solid), in comparison with the 
original flux without the absorption (pink dashed).
}
\label{fig:radfl3}
\end{figure}

Hot bubbles in RGB atmospheres become sources of 
soft X-ray and extreme UV (EUV) radiation that is observed in hybrid stars.   
We present the radiative fluxes, normalized by the bolometric luminosity, from 
the entire star of Models VI as functions of time in Figure \ref{fig:radfl3}.
The blue, green and red lines are the radiative fluxes  
from the gas with $T \le 2\times 10^4\; {\rm K}$ (optical; chromosphere), 
$2\times 10^4 \;{\rm K} < T < 5\times 10^5$ K (UV; transition region) and 
$5\times 10^5\; {\rm K} \le T$ (EUV/soft X-ray; corona).  
We call them cool, warm, and hot components, respectively. 
We also plot the radiative flux of the gas with $T\ge 10^6$ K, which is a 
part of the hot component and corresponds to soft X-ray emission.  

When deriving the radiative fluxes from the entire star, 
we assume spherical symmetry: 
\begin{equation}
L{\footnotesize (T_{1} < T < T_{2})}=4\pi\int dr r^2 
q_{\rm R}{\footnotesize (T_{1}<T<T_{2})}, 
\label{eq:radfl}
\end{equation} 
where $T_1$ and $T_2$ are the minimum and maximum temperatures of 
each component.
While our simulations only treat the open magnetic field regions, in 
reality a certain fraction of the surface is covered by closed structures. 
Equation (\ref{eq:radfl}) assumes that the radiative fluxes from both open 
and closed regions have the same $r$ dependence. 
Although this treatment is not strictly correct, we can give rough 
estimates.  

According to the upper panel of Figure \ref{fig:radfl3}, the hot component 
shows strong intermittent activities, while the cool component, which 
dominates the total radiative flux, stays roughly constant within $10^{-3} 
\lesssim L_{\rm cool}/L_{\star} < 10^{-1}$. The behavior of the warm component 
is between the cool and hot ones; when the hot component is switched-off 
or stays in low activity phases ({\it e.g.} $7500 - 8000$ hr) $L_{\rm warm}/
L_{\star}\approx 10^{-4}$, but once the hot component becomes active,  
$L_{\rm warm}/L_{\star}$ increases to $\gtrsim 10^{-2}$ by the contributions 
from the transition regions of hot bubbles.  

It is also shown that the hot bubbles with $T\ge 10^6$ K 
exist during the only half of the presented duration.  
However, we should note that 
this is based on the spherical symmetry approximation. In reality, 
hot bubbles are created in different open flux tubes at different times, 
and they are ubiquitously distributed 
in the atmosphere to be continuous soft X-ray sources.


When discussing observed X-ray flux, we need to 
take into account the absorption by Hydrogen in the outer cool chromospheric 
wind\citep{ayr05}. Then, the observed flux of the hot component 
would be smaller than that shown in the upper panel of 
Figure \ref{fig:radfl3}.  
In particular, soft X-ray radiation is supposed to severely suffer the 
absorption since most of the hot bubbles with $T>10^6$K are distributed 
in the inner region ($r\lesssim 2 \;R_{\star}$). 

In the lower panel of Figure \ref{fig:radfl3} we compare the observed 
soft X-ray flux ($T\ge 10^6$ K) that takes into 
account the absorption (solid) with the original flux (dashed). 
Tenfold obscuration of the soft X-ray emissions sets in at column 
density, $N_{\rm H} = 5\times 10^{20} - 5\times 10^{21}$ cm$^{-3}$ 
\citep{ayr05}. 
We adopt the most optimistic case 
of $N_{\rm H}=5\times 10^{21}$ cm$^{-3}$: 
\begin{equation}
L = \int dr r^2 q_{\rm R} 10^{-N_{\rm H}/5\times10^{21}\;{\rm cm^{-3}}}.  
\end{equation}
When deriving $N_{\rm H}$ from 
the simulation result, we only consider the gas with $T<8000$ K to 
pick up neutral Hydrogen and assume the solar abundance.
The lower panel of Figure \ref{fig:radfl3} illustrates that most of 
the soft X-ray emissions are absorbed. 
Around $t\simeq 6500$ and 8000 hr, the soft X-ray could be observed, and 
the $L_{\rm X}/L_{\star}$ is reduced to $\lesssim 10^{-8}$ from 
$10^{-5} - 10^{-4}$. Interestingly, this value, $L_{\rm X}/L_{\star} 
\lesssim 10^{-8}$, is typical for hybrid stars (e.g. Ayres 2005), although 
our calculation, which is based on 1D and do not consider detailed 
radiative processes, is too simplified for further quantitative arguments.

\subsubsection{Other RGB Cases}



\begin{figure}
\figurenum{9} 
\epsscale{1.}
\begin{center}
\plotone{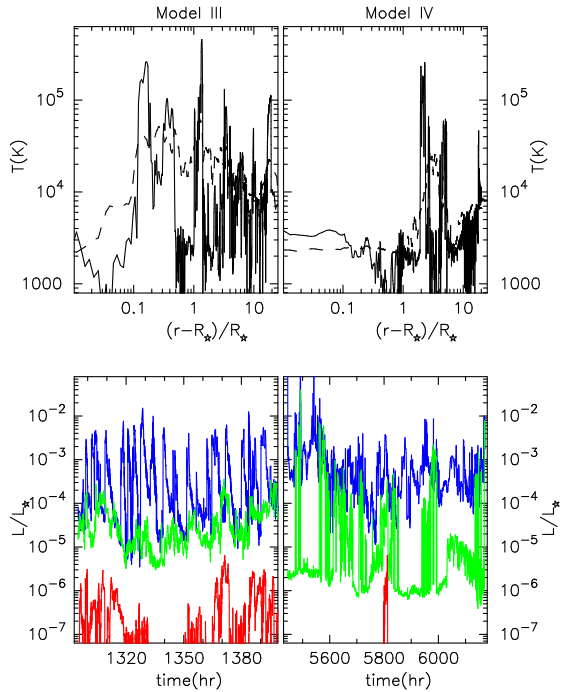}
\end{center}
\caption{
Temperature structure (upper panels)
and time-variations of radiative fluxes (lower panels) 
of Models III (left panels) and IV (right panels). 
The snap-shot results of temperature (solid lines) are at $t=1295$ and 
$6050$hr in Models III and IV, respectively. The time-averaged results 
that are the same as in Figure \ref{fig:wdst}  
are also shown by the dashed lines.   The three components of 
the radiation is the same as in Figure \ref{fig:radfl3}. 
}
\label{fig:tdpd1M}
\end{figure} 


We have examined the time-dependent behaviors of the $3\; 
M_{\odot}$ red giant wind (Model VI). In this subsection we study the 
$1\; M_{\odot}$ stars on both blueward (Model III) and redward (Model IV) 
sides of the DL.  
Figure \ref{fig:tdpd1M} shows the snap-shot temperature structure and 
the time-variation of the radiative fluxes of Models III and IV. 
The upper panels (temperature) show that the atmospheres and winds of 
the $1M_{\odot}$ RGB stars also consist 
of many hot and warm bubbles embedded in the cool chromospheric materials.  
It is  
confirmed that RGB winds are generally structured with many blobs and bubbles. 


The temperature of bubbles are lower in the more evolved star (Model IV).  
This trend is also clearly seen in the plots of the 
radiative fluxes (the lower panels).
The emission 
from the hot gas ($T>5\times 10^5$ K) exists only during the short term 
($t=5800$ hr) in Model VI, while the hot bubbles with $T>5\times 10^5$ K 
almost always exist in Model III. 
Namely, once a star with $1\; M_{\odot}$ 
crosses the DL, 
EUV/soft X-ray emissions from hot plasma with $T>5\times 10^5$ K happen
to disappear.  However, the qualitative wind features, hot/warm 
bubbles in cool materials, are similar in Models III 
and IV. 
Both temperature and the number of hot bubbles decrease with the evolution 
from Models III to IV, because the heating rate per mass decreases 
on account of the increase of the density in the atmosphere 
(Figure \ref{fig:wdst}). 

The emission levels of the hot components of the $1\; M_{\odot}$ stars 
are smaller than that of the $3\; M_{\odot}$ star. 
This is because $\delta v_0$ (Table \ref{tab:sttp}) of the surface convection 
is smaller in a less massive star owing to the lower $T_{\rm eff}$
(Equation \ref{eq:dvscl}). 
Our result is consistent with the observed trend that UV/X-ray emission 
of RGB stars are positively correlated with stellar mass \citep{hs96}.

The average emission level of the hot component of Model III is $L_{\rm hot}
/L_{\star}\approx 10^{-6}$, which is marginally smaller  
than the typical observed X-ray flux ($L_{\rm X}/L_{\star}=10^{-7} - 10^{-4}$; 
e.g. Ayres 2005) from RGB stars blueward of the DL.  
Moreover, the emissions are mostly from the bubbles with $T \lesssim 10^6$(K) 
in our 
simulation, corresponding to EUV rather than soft-X ray emissions, and the 
effect of the absorption will further reduce the observed flux. 
Then, it would be difficult to explain the observed X-ray from RGB stars 
before crossing the DL  
by only the hot bubbles in open field regions.
Probably, hot plasma, which is efficiently confined in closed magnetic loops, 
also contribute to the X-ray emission. 
Incidently, \citet{ros95} speculated that the DL was reflection of the 
change of magnetic topology; as a star crosses the DL, the magnetic structure
becomes open and the hot plasma is liberated to be X-ray deficient 
(see \S\ref{sec:unc}). 
\subsection{Slow Red Giant Winds}
\label{sec:swd}
\begin{figure}
\figurenum{10} 
\epsscale{1.2}
\begin{center}
\plotone{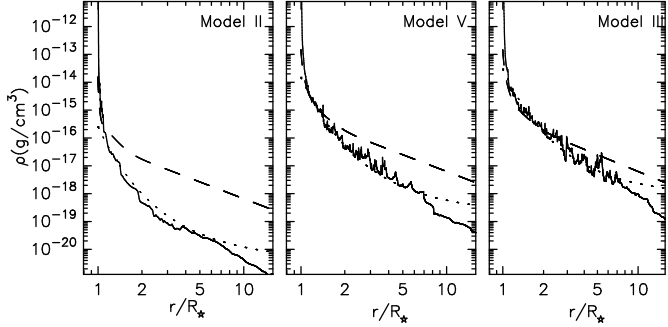}
\end{center}
\caption{Density structure of Model II (left), V (middle), and III (right). 
The results are 
the same as in Figure \ref{fig:wdst} but compared with the densities derived 
from constant wind speeds (dashed lines) and from 
the hydrostatic equilibrium (dotted lines) (see text). 
}
\label{fig:dnst}
\end{figure}

The wind velocity also drastically decreases along with 
the stellar evolution as shown in 
Figure \ref{fig:chp}. 
As briefly stated previously,  
this is mainly because the nearly static regions are formed above 
the photospheres in the RGB stars and 
the winds are essentially accelerated from the effective surfaces located 
several $R_{\star}$. 
We see this in more detail here.  
In order to investigate the drastic change of the wind speed, 
we compare Models 
near the transition at $v_{\rm esc,0}=200-300$ km s$^{-1}$ (the top 
panel of Figure \ref{fig:chp}).
Figure \ref{fig:dnst} shows the simulated density structure (solid) 
of Models II (left), V (middle), and III (right),  
in comparison with the density structure based on hydrostatic equilibrium 
(dotted) and that derived from assumed constant velocities (dashed).

Hydrostatic equilibrium gives  
\begin{eqnarray}
\rho &=& \rho_1 \exp\left(\frac{G M_{\star}}{a_{\rm eff}^2}\left[\frac{1}{r}- 
\frac{1}{r_1}\right]\right) \nonumber
 \\
&=& \rho_1 \exp\left(-\frac{g_1}{a_{\rm eff}^2}\frac{r_{1}}{r}z\right) ,
\label{eq:isdns}
\end{eqnarray}
where $r_1$ is a certain reference position, 
$z \equiv r-r_1$, 
$g_1=G M_{\star}/r_1^2$, and $\rho_1$ is the density at $r=r_1$. 
We here assume the ``isothermal'' atmosphere 
with an effective sound speed, $a_{\rm eff}= a^2 
+ B_{\perp}^2/8\pi\rho$, which takes into account magnetic pressure 
associated with \Alfven waves in addition to 
gas pressure, where $a$ is isothermal sound speed.  
For Models II, V, and III, we adopt $a_{\rm eff} =74, 55$, and 
43 km s$^{-1}$, respectively, which are estimated from the simulation  
results. 
We set $r_1$ at the coronal base (Model II) or the locations where the average 
temperatures start to increase (Models III \& V). Note that 
$r_1\simeq (1.01-1.02) R_{\star}$ in these Models 
(see Figure \ref{fig:chrm} in \S\ref{sec:chm}).  
Figure \ref{fig:dnst} shows that the densities in the subsonic (inner) regions 
can be approximated by the hydrostatic structure,  
which is consistent with the result of general stellar wind theory 
({\it e.g.} section 3 of Lamers \& Cassinelli 1999). 
This indicates that the atmospheres are supported by both magnetic and gas 
pressure. 

The dashed lines plot   
\begin{equation}
\rho \propto (r^2 f)^{-1}, 
\label{eq:ctvl}
\end{equation}
which is derived from substituting constant $v_{r}$ into the steady-state 
continuity equation of mass, $\rho v_r r^2 f=$const.
The starting point is set in the low chromosphere in each Model. 
Because in Model II (the left panel) the density decreases faster due to the 
stronger gravity, the dashed line is far above the solid as 
well as dotted lines. 
This indicates the acceleration of the wind should start near the surface. 

On the other hand, in Models III (the right panel) the decrease of the density 
is slower since the gravity is weaker. As a result, the solid line 
follows the dashed 
line in the inner region, which indicates that the wind is not accelerated, 
but stays almost a constant speed. 
The solid line is gradually diverging from the dashed line around 
$R_{\rm eff} \approx$ several $R_{\star}$, which we call an ``effective'' 
surface, and at which the wind starts to be practically accelerated.
In $r\lesssim R_{\rm eff}$, the average wind speed is kept much slower than 
10 km s$^{-1}$($\ll c_{\rm s}$), and the atmosphere is almost static, 
as seen in Figure \ref{fig:wdst}; the acceleration is not effective 
there because of too much material. 
The longitudinal motions seen in this region in the top panel of 
Figure \ref{fig:wdst} are not by outflow but due to compressive 
waves (\S \ref{sec:dwv}). The wind structure of the other RGB stars 
(Models IV and VI) is essentially similar to that of Model III. 

The gravity is smaller at $r=R_{\rm eff}$ 
than at the surface, hence the slow wind can escape outwardly.    
Therefore, we can expect that the terminal velocity 
is not an order of the surface escape speed
but an order of the escape speed, 
$v_{\rm esc}(R_{\rm eff})$,
at $r=R_{\rm eff}$.  
This can explain why observed wind speeds are much smaller than the 
surface escape velocities in RGB stars ({\it e.g.} Dupree 1986). 
In Model III we can set $R_{\rm eff} 
\simeq 8R_{\star}$, hence, $v_{\rm esc}(R_{\rm eff}) \simeq 70$ km s$^{-1}$.   
This value is an order of the terminal velocity which is supposed to be 
slightly faster than the obtained $v_{\rm 20R_{\star}}=40 - 50$ km s$^{-1}$ 
when the effect of the acceleration in $r>20R_{\star}$ is taken into account.

The situation of Model V (the middle panel) is between Models II and III. 
The simulated density follows the trend of the non-acceleration 
at the beginning. However, the solid line departs from the dashed line 
earlier than in Model III. In this case the wind becomes 
gradually accelerated from a few $R_{\star}$. 
Setting $R_{\rm eff} \approx 3\; R_{\star}$, we have $v_{\rm esc}\approx 
150$ km s$^{-1}$ which is an order of the obtained $v_{\rm 20\; R_{\star}}$. 
The difference between Models III and V arises from the difference 
of the radii;  
although these two Models have the same $g$, 
the decrease of the density is faster in Model V when measured in 
$r/R_{\star}$,  ($z=r-R_{\star}$ is larger in Equation \ref{eq:isdns}).


In concluding, we have the natural conclusion that the wind speed is 
determined by the escape speed at $r=R_{\rm eff}$ where the wind practically 
starts to flow out. 
$R_{\rm eff}$ does not coincide with the stellar surface but moves 
outward as a star expands. 
Therefore, the wind speeds of the RGB stars 
are slower than the surface escape speeds. 



\subsection{Energetics}
The main energy that drives the stellar winds is the turbulent energy of 
the surface convection in our simulations. The \Alfven waves that are 
excited by the footpoint motions of open flux tubes carry the energy outwardly 
and the dissipation of the waves leads to the heating and acceleration of  
the stellar winds.  In this subsection, we quantitatively investigate 
the energetics of the stellar winds driven by \Alfven waves, and 
finally determine the mass loss rate in a {\it forward} manner. 

\subsubsection{Heating}
\label{sec:chm}

\begin{figure}
\figurenum{11} 
\epsscale{1.}
\begin{center}
\plotone{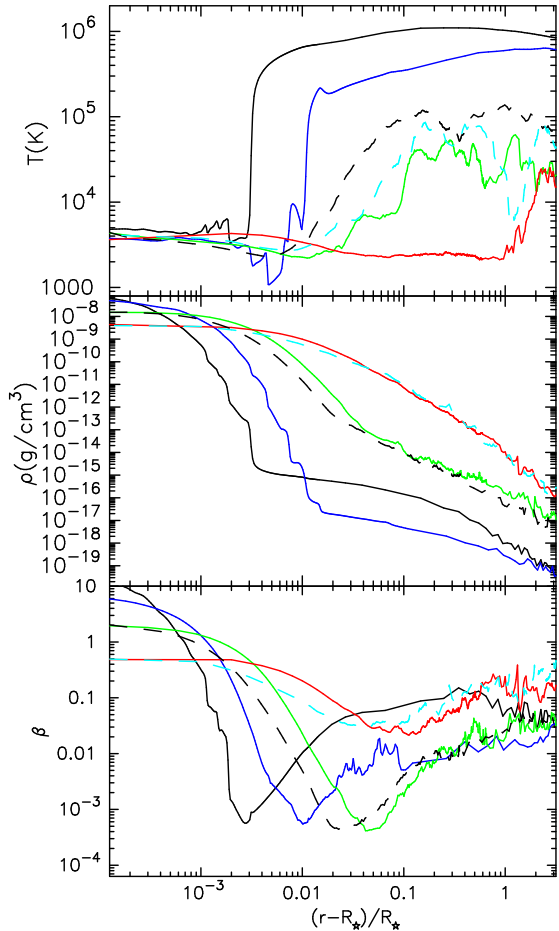}
\end{center}
\caption{Structure of the inner regions of the six Models. 
From the top to the bottom, $T$ (K), $\rho$ (g cm$^{-3}$), and plasma 
$\beta$ values are shown on
$(r-R_{\star})/R_{\star}$. 
The results in the top and middle panels are the same as in Figure 
\ref{fig:wdst}.
The solid lines are the results 
of the 1 $M_{\odot}$ stars; the black, blue, green, and red lines are the 
results of Models I, II, III, and IV, respectively, 
The dashed lines are the results of the $3\; M_{\odot}$ stars; the black and 
light-blue lines are the results of Models V and VI. }
\label{fig:chrm}
\end{figure}

We study the structure of lower atmosphere because it is important not only 
in determining the basal properties of the stellar winds but in controling the 
reflection of the \Alfven waves.
In Figure \ref{fig:chrm} we show the time-averaged density (top) and 
temperature (middle) 
which are the same as in Figure \ref{fig:wdst} but plotted 
on $(r-R_\star)/R_\star$ to enlarge the inner regions.  
Note that the physical distance ({\it e.g.} km) of the horizontal axis 
becomes larger for a star with larger $R_{\star}$. 
In the bottom panel plasma $\beta$ values defined as
\begin{equation}
\beta \equiv \frac{8\pi p}{B^2}
\end{equation} 
are plotted. 

In the MS (Model I) and sub-giant (Model II) cases, the steady corona lies 
above the (geometrically) thin chromosphere and these two regions 
are divided by the sharp transition region. In the RGB cases (Models 
III-VI), the temperature gradually increases outwardly from the chromosphere. 
This increase of the temperature is different from the formation of ``warm 
corona'' but the result of the time-average of intermittent creation 
of hot bubbles; the location at which the temperature increases is 
the starting point of the formation of hot bubbles.     
However, it is still worth studying the average trend from an energetics point 
of view. 

The lower panel illustrates that the variations of $\beta$'s are qualitatively 
similar in the stars with $\log g=2.4 - 4.4$; 
the $\beta$ values decrease outwardly to $\sim 10^{-3}$ but turn to 
increase from certain locations.  
This is because when the magnetic pressure dominates 
the gas pressure by $\sim$1000 times, the dissipation of a small fraction 
of magnetic energy gives an enormous effect on the gas. 
Indeed, the energy transfer from the magnetic fields to the plasmas is 
carried by the dissipation of the \Alfven waves in our simulations and the 
gas is heated up before $\beta$ becomes further smaller. 
As the temperatures rise, the decreases of the densities become slower 
(Equation \ref{eq:isdns}) and $\beta$'s increase due to  
the decreases of $B$'s. 
The densities at which the average temperatures start to increase are 
controlled in order that the plasma $\beta$ values do not become too small.  
The geometrical depth of the inner chromosphere (the region between the 
photosphere and the location of temperature rise) is larger in a lower 
$\log g$ star since the decrease of density is slower. 

In the stars with $\log g=1.4$, the $\beta$ values 
do not become as small as the other cases since the decreases of the densities 
in the low chromospheres are too slow to achieve such small $\beta$'s. 
Therefore, strong heating do not occur and the average 
temperatures are lower than the higher gravity stars. 


In the $3M_{\odot}$ stars the temperatures start to increas from more 
inner positions than in the $1M_{\odot}$ counterparts. 
This is because 
a more massive star gives larger photospheric amplitude and input 
energy of \Alfven waves (Equation \ref{eq:dvscl}). 
Then, hot bubbles are heated up from a deeper ({\it i.e.} denser) region. 

\subsubsection{Dissipation and Reflection of Waves}
\label{sec:dwv}
In SI05 and SI06, we claimed that the nonlinear generation of MHD slow (
$\approx$ sound) waves is the main dissipation process in our 1D simulation 
of the solar wind; the energy of \Alfven waves is transferred to the 
slow waves, which dissipate by shocks after steepening of the wave fronts. 
We show that the same mechanism operates in RGB stars at least under the 
1D approximation.
We firstly inspect the dissipation processes by using the results of Model 
III, and after that, we compare 
the four cases of the $1\; M_{\odot}$ stars (Models I-IV).  

\begin{figure}
\figurenum{12} 
\epsscale{0.9}
\begin{center}
\plotone{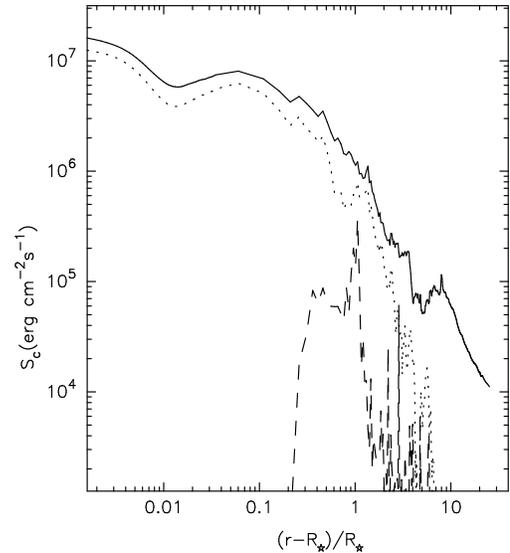}
\end{center}
\caption{Adiabatic constants, $S_{\rm c}$'s (erg cm$^{-2}$s$^{-1}$), of 
outgoing \Alfven (solid), incoming \Alfven(dotted), and outgoing slow (dashed) 
MHD waves in Model III.    
}
\label{fig:wac}
\end{figure}

We are injecting the broadband spectra of the surface fluctuations and   
in the upper regions not only outgoing but incoming waves of different modes 
travel simultaneously. 
Hence, the actual simulations show turbulence-like feature 
({\it e.g.} see $v_{\perp}$ in the movie of Figure \ref{fig:mpw} available in 
{\it electric edition}). 
A technic of the mode decomposition \citep{cl03} is a useful tool to study 
such turbulent phenomena.    
Figure \ref{fig:wac} plots the following adiabatic constants derived from 
wave action \citep{jaq77},   
\begin{equation}
\mbf{S}_c
=\rho \langle \delta v_{\rm w}^2 \rangle \frac{(v_r + v_{\rm ph})
(\mbf{v_r + v_{\rm ph}})}{v_{\rm ph}} \frac{r^2 f(r)}{r_c^2 f(r_c)}, 
\end{equation} 
for outgoing Alfv\'{e}n, incoming Alfv\'{e}n, and outgoing slow MHD (sound) 
waves of Model III as functions of distance, 
where $\delta v_{\rm w}$ and $\mbf{v_{\rm ph}}$ are amplitude and phase 
speed of each wave mode. 
In our simple 1D geometry, \Alfven waves are purely transverse and slow MHD 
waves are purely longitudinal; the energy flux of \Alfven waves is calculated 
from $v_{\perp}$ and $B_{\perp}$, and the energy flux of slow MHD waves 
is from fluctuating components of  $v_r$ and $\rho$.  
A characteristic feature of incoming (outgoing) \Alfven waves is that 
$v_{\perp}$ and $B_{\perp}$ are in (in opposite) phase. 
The decomposition of each wave mode can be done by using these facts\footnote{
This method is strictly valid only for 
linear and short wavelength waves of which amplitudes are sufficiently 
smaller than phase speed and wavelengths are sufficiently shorter than 
a variation scale of background phase speed ({\it i.e.} WKB approximation 
holds).
For nonlinear and long wavelength waves, which we want to deal with, this 
treatment is only {\it statistically} true \citep{cl03}. }.
The subscript, $c$, denotes that $\mbf{S}$ is normalized at 
$r_c$ where the flux tubes expand by $f_c=40$, to exclude the effect of 
the adiabatic loss. 

$\mbf{S}_c$ can be used as a measure of the dissipation. Although 
$\mbf{S}_c$ and energy flux are identical in static media, they are different 
in moving media. 
$\mbf{S}_c$ is conserved in expanding atmosphere if a wave does 
not dissipate, while it is not the case for the energy flux. 
For the incoming \Alfven wave, we plot the opposite sign of 
$\mbf{S}_c$ so that it becomes positive in the sub-\Alfvenic region.

Figure \ref{fig:wac} shows that the outgoing \Alfven waves dissipate quite 
effectively;   
$S_c(=|\mbf{S}_c|)$ becomes only $\sim 10^{-3}$ of the initial value 
at the outer boundary. 
First, a sizable amount is reflected back quite near the surface, 
which is clearly illustrated as the incoming \Alfven wave that is following 
the outgoing component with a slightly smaller level.  
This is because the wave shape is considerably 
deformed owing to the steep density gradient. In this region, the temperature 
is small, $T\simeq 3000$ K, which gives a short pressure scale height, 
$H_{\rm P}$ (Figure \ref{fig:chrm}). 
Then, the wavelength of 
the \Alfven waves 
becomes comparable or longer than $H_{\rm P}$, which results in 
the effective reflection of the waves \citep{moo91}. 
In this Model III, more than 60\% 
of the initial energy flux of the \Alfven waves are reflected back before 
reaching the upper warm region with the {\it average} temperature, 
$\sim 10^5$ K (\S\ref{sec:eng}).

Second, slow MHD waves are generated in the corona (see Sakurai et al.2002 
for solar observation) as shown in Figure \ref{fig:wac}. 
The generation of the slow waves is triggered by the variation of the 
magnetic pressure, 
$B_{\perp}^2/8\pi$, with the \Alfven waves \citep{ks99}.  
This excites density perturbations, which are slow MHD waves in magnetically 
dominated plasma. Slow waves are generally longitudinal, 
and then, they eventually suffer 
nonlinear 
steepening to lead to shock dissipation \citep{suz02}.  
The density fluctuations of slow waves also work as mirrors to 
\Alfven waves. The variation of the density directly leads to the variation of 
the \Alfven speeds ($\propto B/\sqrt{\rho}$), which enhances the reflection 
of the \Alfven waves by the deformation of the shape. These reflected waves 
further nonlinearly interact with the pre-existing outgoing waves, which 
plays a role in the dissipation.   

\begin{figure}
\figurenum{13} 
\epsscale{1.}
\begin{center}
\plotone{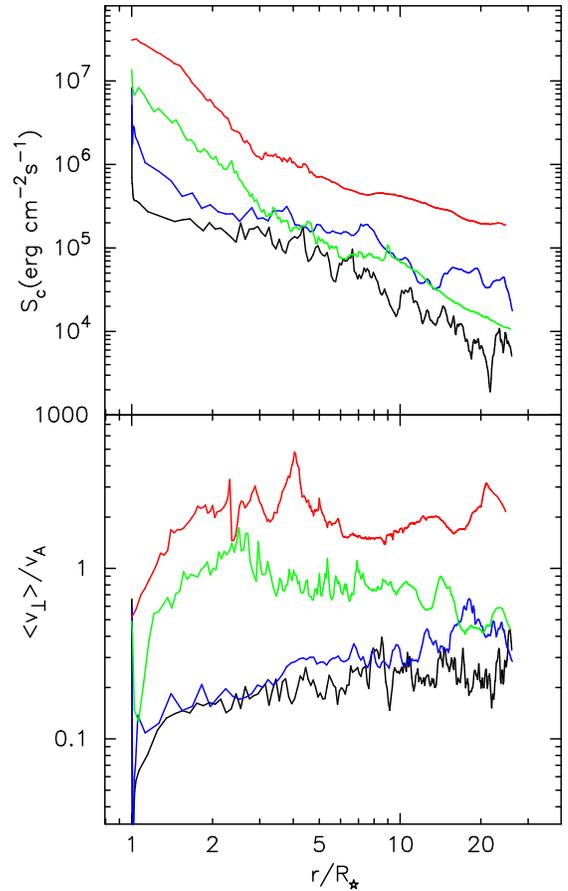}
\end{center}
\caption{Properties of \Alfven waves in the 1 $M_{\odot}$ stars 
(Models I - IV). The top panel exhibits the adiabatic constant, 
$S_{\rm c}$, and the bottom panel shows the nonlinearity, 
$\langle v_{\perp}\rangle/v_{\rm A}$. 
The black, blue, green, and red lines are the 
results of Models I, II, III, and IV, respectively,
}
\label{fig:wvcmp}
\end{figure}

The process discussed here is the parametric decay instability of \Alfven 
waves due to three-wave (outgoing Alfv\'{e}n, incoming Alfv\'{e}n, 
and outgoing slow waves) interactions (Goldstein 1978; Terasawa et al. 1986). 
The decay instability is not generally efficient in the homogeneous background 
since it is a nonlinear mechanism. Therefore, it has not been considered 
as a major process of the acceleration of the solar and stellar winds.  
However, the density gradient of the 
background plasma totally changes the situation. 
The amplitude, $v_{\perp}$, of \Alfven waves is inevitably amplified 
with the decreasing density to conserve $S$. On the other hand, 
the \Alfven speed, $v_{\rm A}$($\propto B/\sqrt{\rho}$), does not increase 
so much because the decrease of $B_r$ (flux tube expansion) compensates 
the decrease of $\rho$.  
As a result, the 
nonlinearity ($\langle v_{\perp}\rangle/v_{\rm A}$) increases as the 
\Alfven waves travel 
upwardly (see Figure \ref{fig:wvcmp}), which leads to efficient dissipation.  

Figure \ref{fig:wvcmp} compares the dissipation of the \Alfven waves in the 
four Models of the 1 $M_{\odot}$ stars. In the upper panel $S_c$'s are 
compared, 
and in the lower panel nonlinearities, $\langle v_{\perp}\rangle /v_{\rm A}$, 
are shown, where $\langle v_{\perp}\rangle=\sqrt{\langle v_{\perp}^2\rangle}$. 
The figure shows that the wave energies drastically attenuate just above the 
surfaces in higher gravity stars because of the reflection in 
the chromospheres.  In these stars the densities decrease more rapidly in the  
inner chromospheres, leading to the increases of $v_{\rm A}$'s. 
Then, the wavelength of the \Alfven wave with the typical frequency 
(e.g. several min. 
in Model I) becomes longer than $H_p \propto 1/g$ which itself is 
substantially smaller for larger $g$. This enhances 
the reflection as explained above. 

Furthermore, in stronger gravity stars, the 
chromospheres are 
geometrically thin ($r-R_{\star} \ll R_{\star}$). Otherwise if the
chromosphere extends to $r-R_{\star} \gtrsim R_{\star}$ as in the RGB stars, 
a curvature effect increases $H_p$ by $r/R_{\star}$ (Equation \ref{eq:isdns}), 
which reduces the wave reflection.   
This is also a reason why the reflection becomes relatively important in 
the higher gravity stars. 
Finally, only less than half of the input wave energies can go beyond 
the inner chromospheres in Models I - III (\S\ref{sec:eng}). 
     
Let us turn to the waves in the outer regions. Wave dissipation is more rapid  
in lower gravity stars. 
This is because the densities in the atmopsheres are larger in those 
stars due to the longer $H_p$
and accordingly the \Alfven speeds, 
$\propto 1/\sqrt{\rho}$, becomes slower. In such a condition \Alfven waves 
more easily become nonliear ($\langle v_{\perp}\rangle /v_{\rm A}\gtrsim 1$ 
in the bottom 
panel of Figure \ref{fig:wvcmp}), which enhances the wave dissipation.      
Then, one may expect that the wave energies are more effectively transferred 
to the winds in more evolved stars. 
However, the situation is 
not so simple because in these stars radiative losses become efficient due to 
the higher densities. We study the wind energetics to discuss these issues 
in the next subsection.  

Before closing this subsetion we would like to remark on dissipation 
lenghts of the \Alfven waves.  
Since the dissipation processes of \Alfven waves are quite complicated, 
a simpliphied phenomenological approach has been often adopted 
to treat \Alfven waves in solar and stellar wind models ({\it e.g.} Hartmann 
\& MacGregor 1980);  
an action of \Alfven waves is assumed to follow 
\begin{equation}
S \propto \exp\left(-\frac{r-R_{\star}}{l} \right), 
\end{equation} 
where $l$ is a (constant) dissipation length ($=0.1 - 10\; R_{\star}$). 
We can directly estimate $l$ from our simulations. The upper panel of 
Figure \ref{fig:wvcmp} illustrates that the decreases of $S_c$'s are roughly 
on straight lines in the $\log(r/R_{\star})- \log S_c$ diagram. This means 
that the dissipation lengths increase on $r$. For example, 
in Model IV $l \lesssim 1\; R_{\star}$ in $r<3\; R_{\star}$, but 
$l \approx 10\; R_{\star}$ in $r>10\; R_{\star}$;  
the assumption of a constant dissipation length is not 
appropriate.

\subsubsection{Energy Transfer from \Alfven Waves}
\label{sec:eng}

\begin{figure}
\figurenum{14} 
\epsscale{1.}
\begin{center}
\plotone{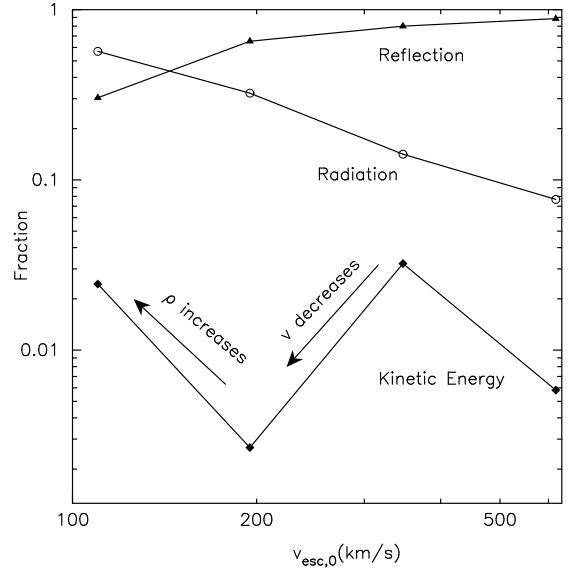}
\end{center}
\caption{The fractions of wind kinetic energy, radiative loss, 
and wave energy reflected back to the surfaces of the 1 $M_{\odot}$ 
stars converted from the input energy of \Alfven wave 
as functions of $v_{\rm esc,0}$ (km s$^{-1}$). 
}
\label{fig:engfrc}
\end{figure}

A fraction of the initial wave energy is used for 
the acceleration of the stellar winds, while 
the rest of them is lost by radiation and returned back to the 
stellar surface by counter propagating waves. 
We estimate kinetic energy fluxes of the stellar winds at the outer 
boundaries, 
integrated radiative losses, and wave energy fluxes leaking inward from the 
stellar surfaces, respectively, in unit of erg/s :
\begin{equation}
(\rho v_r \frac{v_r^2}{2} f r^2)_{\rm out},
\end{equation}   
\begin{equation}
\int_{R_{\star}}^{r_{\rm out}}dr r^2 f q_{\rm R},
\end{equation}
\begin{equation}
(\rho \langle \delta v_{\perp,-} ^2 \rangle v_{\rm A})_{\star} R_{\star}^2,  
\end{equation}
where $\delta v_{\perp,-}$ denotes amplitude of incoming \Alfven waves, and 
subscripts `$\star$' and `out' indicate that the values are evaluated at 
the stellar surfaces and the outer boundaries, respectively.     
Here, we do not discuss energy carried by waves, enthalpy, and thermal 
conduction away from the outer boundaries because 
their contributions are smaller. 
Figure \ref{fig:engfrc} presents these energies normalized by the input energy 
fluxes from the surfaces, 
\begin{equation}
(\rho \langle \delta v_0^2 \rangle v_{\rm A})_{\star} R_{\star}^2 , 
\end{equation}
as functions of $v_{\rm esc,0}$.

As one can see, the fractions of the wave energies that are converted to the 
wind kinetic energies are small. Typically, only $\sim$ 1 \% of the surface 
turbulent energies are transferred to the stellar winds. However, these 
tiny fractions are enough to keep the observed $\dot{M}$ and 
$v_{\infty}\approx v_{20R_{\star}}$ of the winds from MS and RGB stars. 
  
Most of the energy is lost by the radiation and the wave reflection. 
In the less evolved stars (Models I-III), the energy losses are 
dominated by the wave 
reflection as discussed in \S\ref{sec:dwv}; the densities decrease more 
rapidly because of 
the higher gravity, and then, 
the wavelengths become longer than the scale heights, which results in the 
effective reflection via the deformation of the wave shape.   
The rapid decreases of the 
densities also make the radiative cooling less effective; it is important 
only in the chromospheres ($T<10^4$ K) and low coronae where the densities are 
sufficiently high (volumetric radiative cooling is $\propto 
\rho^2$ in optically thin and $\propto \rho$ in optically thick gas). This 
also leads to the relative dominance of the wave reflection in the total 
energy loss.  
On the other hand, the trend of the most evolved star (Model IV) is opposite 
since the decrease of the density is more gradual; 
radiative cooling becomes relatively important first because it is efficient 
even in the outer region and second because the \Alfven waves do not suffer 
the reflection so much.

Let us turn back to the kinetic energy part. The tendency on $v_{\rm esc,0}$ 
is not monochromatic. From Model I (MS; $v_{\rm esc,0}=617$ km s$^{-1}$) 
to II (sub-giant; $v_{\rm esc,0}=347$ km s$^{-1}$), the 
fraction of the kinetic energy increases mainly because the mass flux, 
$\rho v_r$,  
increases. However, the fraction drops almost 
by one order of magnitude from Models II to III 
($v_{\rm esc,0}=195$ km s$^{-1}$). This is because of the 
drastic decrease of the wind speed (Figure \ref{fig:chp}). Although the 
density increases from Models II to III, the effect of the velocity dominates 
since the kinetic energy flux is proportional to $\rho v_r^3 $. 
From Models III to IV ($v_{\rm esc,0}=110$ km s$^{-1}$), the decrease of $v_r$ 
is small, and the fraction of the kinetic energy again increases simply by
the increase of the density.

\subsubsection{Mass Loss Rate}
\label{sec:msls}
The bottom panel of Figure \ref{fig:chp} shows the mass loss rate of 
the $1M_{\odot}$ star rapidly 
increases by $\sim 5$ orders of magnitude from MS to RGB phases. 
However, the increase of $\dot{M}$ is continuous, which is in contrast to 
the velocity and temperature showing the gaps between the sub-giant and 
RGB stars. 
This is because the density in the wind drastically rises during this phase, 
which compensate the rapid decrease of the wind speed.

\begin{figure}
\figurenum{15} 
\epsscale{1.}
\begin{center}
\plotone{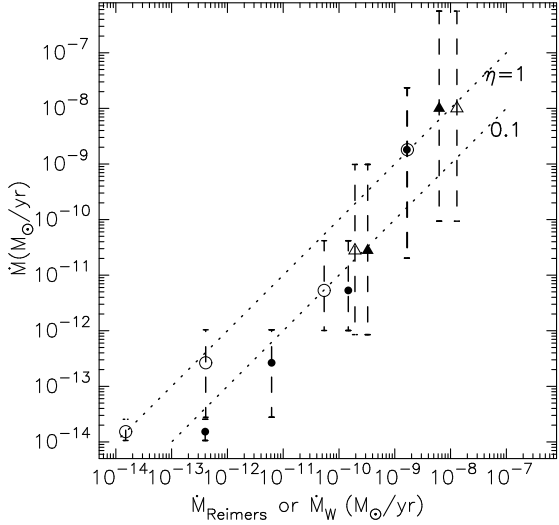}
\end{center}  
\caption{Comparison of the mass loss rates of the simulations 
(vertical axis) with the Reimers' relation 
(filled symbols; horizontal axis) and 
Equation (\ref{eq:mswv}) derived from the scaling relations of wind energetics 
(open symbols; horizontal axis). The circles and the triangles indicate the 
results of the $1M_{\odot}$ and $3M_{\odot}$ stars, respectively.
}
\label{fig:reim}
\end{figure}

Since our simulations adopt a {\it forward} approach, $\dot{M}$ is not 
an input parameter but an output directly determined from the simulations. 
Therefore, it is straightforward to study how $\dot{M}$ evolves with 
the stellar evolution from the wind energetics.  
We begin by comparing our results with the empirical relation  
by \citet{rei75}:
\begin{equation}
\dot{M}_{\rm Reimers} = 4\times 10^{-13} \eta_{\rm R} 
\frac{(L_{\star}/L_{\odot})(R_{\star}/R_{\odot}) }
{({M_{\star}/M_{\odot}})}\;\; M_{\odot}/{\rm yr},  
\label{eq:rei}
\end{equation}
where $\eta_{\rm R}$ is a proportional coefficient of an order of unity.  
This is based on a dimensional analysis of stellar wind energetics. 
The filled symbols in Figure \ref{fig:reim} compares the obtained 
$\dot{M}$ with the $\dot{M}_{\rm Reimers}$ 
for $\eta_{\rm R} =1$. 
The comparison indicates that the Reimers' relation roughly gives reasonable 
estimates. 
However, smaller $\eta_{\rm R} (\lesssim 0.1)$ seems to be systematically 
favored in the higher gravity stars.   
To search a better relation, we investigate the mechanism that controls 
$\dot{M}$ from the energetics of our Alfv\'{e}n-wave driven wind simulations. 

In the wave-driven winds, the radiation part ($L_{\star}$) should 
be replaced by wave energy in Equation (\ref{eq:rei}).  
In an open flux tube, a fraction, $\epsilon$, of the initial energy flux of 
input \Alfven wave is transferred to the stellar wind where $\epsilon$ is 
exactly what has been derived in \S \ref{sec:eng} (Figure \ref{fig:engfrc}). 
This can be written as 
\begin{equation}
\epsilon R_{\star}^2 F_{\rm w,\star} = r^2 f \rho v_r \frac{v_r^2}{2},   
\label{eq:wvwde}
\end{equation} 
where $F_{\rm w,star}=\rho_0 \langle \delta v_0^2 \rangle v_{\rm A}$ is wave 
energy flux. The left-hand side is evaluated at the surface and the 
right-hand side is at a sufficiently distant location at which 
$f=f_{\rm tot}$ and $v\approx v_{\infty}$. 
Multiplied by $4\pi$, a transformation of Equation (\ref{eq:wvwde}) gives 
\begin{equation}
\epsilon L_{\rm w}=\dot{M}\frac{v_{\infty}^2}{2}, 
\end{equation}
where $L_{\rm w}=4\pi R_{\star}^2F_{\rm w}/f_{\rm tot}$ is wave luminosity 
and $\dot{M}=4\pi \rho v_{\infty} r^2$.  
Here $F_{\rm w}/f_{\rm tot}$ is used instead of $F_{\rm w}$ because 
the fraction of the area occupied by open field regions at the surface is 
$1/f_{\rm tot}$. 
Then, an equation that determines $\dot{M}$ becomes 
\begin{equation}
\dot{M_{\rm w}}=2\epsilon\frac{L_{\rm w}}{v_{\infty}^2} 
=\left(2\epsilon\frac{v_{\rm esc,0}^2}{v_{\infty}^2}
\frac{L_{\rm w}}{L_{\star}}\right)\frac{L_{\star}}{v_{\rm esc,0}^2} 
=\left(\epsilon\frac{v_{\rm esc,0}^2}{v_{\infty}^2}
\frac{L_{\rm w}}{L_{\star}}\right) \frac{L_{\star}R_{\star}}{G M_{\star}},
\label{eq:mslsd}
\end{equation}
where we add subscript 'w' to distinguish from $\dot{M}_{\rm Reimers}$ and 
the simulated $\dot{M}$. 
Although $\epsilon$ varies on $v_{\rm esc,0}$, we have almost constant 
\begin{equation}
\epsilon{v_{\rm esc,0}^2}/{v_{\infty}^2} \approx 0.02, 
\label{eq:eps}
\end{equation}
because $v_{\infty}^2/{v_{\rm esc,0}^2}$ is positively correlated with 
$\epsilon$ as shown in Figure \ref{fig:engfrc}.    

As for the wave luminosity, we can use the scaling relations of 
Equations (\ref{eq:sfprs}) and (\ref{eq:dvscl}), and then, 
\begin{eqnarray}
\frac{L_{\rm w}}{L_{\star}}=\frac{F_{\rm w}/f_{\rm tot}}{\sigma 
T_{\rm eff}^4} &=& \frac{(B_{r,0}/f_{\rm tot})}{\sqrt{4\pi}\sigma}
\frac{\sqrt{\rho_0} \langle \delta v_0^2\rangle}
{T_{\rm eff}^4} \nonumber \\
&=& 1.4\times 10^{-5}\frac{(T_{\rm eff}/T_{\rm eff,\odot})^9}
{(g/g_{\odot})} 
\label{eq:lrt}
\end{eqnarray}
where $\sigma$ is the Stephan Boltzmann constant, subscript, $\odot$, denotes 
the solar values, and we are using the 
same $B_{r,0}/f_{\rm tot}$($=1$ G) in all the Models.  

Substituting Equations (\ref{eq:eps}) and (\ref{eq:lrt}) into 
Equation (\ref{eq:mslsd}), we finally have 
\begin{eqnarray}
\hspace{-0.5cm}\dot{M_{\rm w}} &=& 1.7\times 10^{-14}\eta_{\rm w}
\frac{(L_{\star}/L_{\odot})(R_{\star}/R_{\odot})
}{M_{\star}/M_{\odot}}\frac{(T_{\rm eff}/T_{\rm eff,\odot})^9}{g/g_{\odot}}
\label{eq:mswv1}
\\
&=& 
1.7\times 10^{-14}\eta_{\rm w}
\frac{(L_{\star}/L_{\odot})(R_{\star}/R_{\odot})^3
(T_{\rm eff}/T_{\rm eff,\odot})^9}{(M_{\star}/M_{\odot})^2}
\label{eq:mswv}
\end{eqnarray}
in $M_{\odot}$yr$^{-1}$, where we leave a degree of freedom in 
$\eta_{\rm w}(\approx 1)$, similarly to the Reimers' relation 
(Equation \ref{eq:rei}) since $\epsilon$ is not a constant. 
Note that the dependences of $g$ and $T_{\rm eff}$ in Equation 
(\ref{eq:mswv1}) are from the dependences of the energy flux of 
the generated sound waves by \citet{ren77} (Equation \ref{eq:acfsc});  
if we used other results by {\it e.g.} \citet{ste67,boh84}, which give 
slightly different dependences, the scaling relations of $\dot{M_{\rm w}}$ 
would be also slightly different. 
$\dot{M}_{\rm w}$ does not have dependence on the properties of 
magnetic fields since we assume a constant $B_{r,0}/f_{\rm tot}$($=1$ G) 
in Equation (\ref{eq:lrt}). 
In more realistic situations, however, $B_{r,0}/f_{\rm tot}$ would show 
star-to-star variation, and $\dot{M}_{\rm w}$ would have dependence on 
the magnetic fields.

The open symbols in Figure \ref{fig:reim} compares the simulated 
$\dot{M}$ and $\dot{M}_{\rm w}$. One can see that most of the data points 
are distributed near $\eta_{\rm w}=1$. 
While the Reimers' formula, mainly targeting more luminous 
red giants, does not give good $\dot{M}$ estimates for the less evolved stars 
(filled symbol; Equation \ref{eq:rei}), our 
scaling gives a better explanation to these stars. 
Then, the scaling of Equation (\ref{eq:mswv}) can 
connect the mass loss rate of less evolved stars to that of more evolved ones 
that can be handled by the Reimers' formula.

\citet{sc05} also proposed a formula of mass loss rates of wave-drive winds, 
focusing on more evolved stars. 
In their formula, additional dependences on $g$ and $T_{\rm eff}$ also appear, 
while they are different from ours. 


\section{Discussion}
\label{sec:dis}
\subsection{Magnetic Fields}
\label{sec:unc}
When performing the simulations, we have to set up in advance the surface 
fluctuations and 
the magnetic flux tubes.  
While we can reasonably estimate the amplitude and period (spectrum) 
of the surface fluctuations from the conditions of the convection zones 
(\S \ref{sec:phfl}),  
there are large uncertainties in the properties of the magnetic fields.

We only consider the open magnetic field regions because our aim is to study 
the stellar winds. 
We assume the same magnetic field strength ($B_{r,0}=240$ G) at the surface 
and expansion factor ($f_{\rm tot}=240$) in the different stars.  
Although it is unclear how in reality the open magnetic flux tubes are,
we had better consider effects of different flux tubes 
(see \citet{fvj06} for configurations of flux tubes). 

We have actually studied the effects of the strength and configuration of 
magnetic flux tubes on the solar wind by both numerical simulations 
(SI06) and analytical method \citep{suz06}.  
We showed that $B_{r,0}/f_{\rm tot}$ is a more important parameter in 
controling solar wind properties than individual $B_{r,0}$ and $f_{\rm tot}$, 
if nonlinear 
\Alfven waves dominantly work in the acceleration. This is also consistent 
with recent observation by \citet{kfh05}. 
For example, flux tubes with different $B_{r,0}$ and $f_{\rm tot}$ but the 
same ratio $B_{r,0}/f_{\rm tot}$ give similar wind properties if the 
photospheric amplitudes are the same. 
An input energy flux is $\sim \rho \langle \delta v_0 \rangle^2 v_{A,0} 
\propto B_{r,0}$ at the surface, and it is diluted according to 
$1/fr^2$ in an expanding 
flux tube. The acceleration of the stellar wind mainly occurs in the outer 
region where the super-radial expansion of the flux tube already finishes. 
Therefore, 
$B_{r,0}/f_{\rm tot}$, which is proportional to $S_c$ ($\approx$energy flux) 
of the input \Alfven wave {\it measured at the outer region}, becomes an 
important control parameter in the \Alfven wave-driven wind. 
Moreover, a mass flux, {\it i.e.} $\dot{M}$, is determined almost solely by 
$\langle \delta v_0 \rangle^2$, 
and very weakly dependent on $B_{r,0}/f_{\rm tot}$ (SI06). 
Then, $B_{r,0}/f_{\rm tot}$ plays a role only in 
tuning the terminal velocity of the solar wind. 

This argument is directly applicable to red giant winds. 
$\dot{M}$'s of the stellar winds (the top panel of 
Figure \ref{fig:chp}) are determined with little uncertainty since they 
are mainly determined by $\langle \delta v_0 \rangle^2$ 
of the surface convections. 
However, when we discuss the wind terminal speeds, $B_{r,0}/f_{\rm tot}$ needs 
to be taken carefully; wind speed becomes faster in a flux tube with larger 
$B_{r,0}/f_{\rm tot}$. 

$B_{r,0}/f_{\rm tot}(=240/240=1\; {\rm G})$ corresponds to the total 
magnetic flux over the whole open field regions divided by the 
entire stellar surface. 
To discuss the average field strength we have 
to take into account closed regions, which occupy a sizable fraction of 
the surface. If 10\% of the surface is occupied by open field 
regions, the average strength becomes 10 times, namely 10G. 
This value is quite reasonable from an energetics point of view. 
The magnetic energy density at the surface is 
$B_{r,0}^2/8\pi \simeq 4$ erg cm$^{-3}$, 
which is considerably smaller than the kinetic energy of the surface 
turbulence, $\rho \langle \delta v_0^2 \rangle \sim 10^{3} - 10^4$
erg cm$^{-3}$, of the simulated stars.
It is sufficient only if $\sim 0.1$ \% of the turbulent energy is 
converted to the magnetic energy by motions of charged particles ({\it i.e.} 
current) in the surface convection. 

Related to this, \citet{ros91,ros95} tried to interpret the DL 
by the change of magnetic topologies. 
The stellar rotation becomes slow with the evolution 
so that the generation of magnetic fields by 
large-scale dynamo is switched-off. Magnetic fields are generated only by
(small-scale) turbulent convection in the surface layers in evolved 
stars.   
As a result, 
closed fields, which have dominantly covered the surface in unevolved stars, 
become open in evolved stars redward of the DL. 
Therefore, hot plasma, which has been well 
confined by closed loops, would flow out.

If this scenario is correct\footnote{It should be noted that \citet{ahb03} 
recently proposed a different scenario for the DL: even though closed loops, 
which are sources of X-rays, are distributed in stars redward of the DL, 
these X-rays 
are obscured by larger column density of above chromospheric material. 
This density effect is a reasonable explanation of the DL.}, 
surface field strength, $B_{r,0}$, becomes smaller in evolved stars. 
However, we expect that the field strength in the outer regions 
($\propto B_{r,0}/f_{\rm tot}$) is less affected, because an decrease 
of $B_{r,0}$ firstly reduces a fraction of closed structures rather than 
strength of open fields in the outer region;  
even if $B_{r,0}$ decreases, $B_0/f_{\rm tot}$ stays more or less constant 
before all the field lines become open. 
Because $B_{r,0}/f_{\rm tot}$ is more important 
than $B_{r,0}$ as stated above, we anticipate that our results of the 
stellar winds 
are affected little even if the magnetic field topologies change with
the stellar evolution.

\subsection{Previous Wind Calculations} 
\citet{hm80} investigated stellar winds from late-type stars with 
explicitly taking into account the effects of \Alfven waves. 
They calculated several stars with different wave 
energies and dissipation parameters under the steady-state condition 
from assumed inner boundaries. 
They concluded that the dissipation of \Alfven waves resulted in 
coronal heating in higher gravity stars (MS and sub-giant) and mass loss 
in lower gravity stars (RGB). 
Our analysis of the time-average wind structures basically confirm this 
tendency, 
whereas our simulations have revealed the various 
important time-dependent aspects (\S\ref{sec:tdp}).      

Most of previous works claim that it is difficult to accelerate slow and dense 
red giant winds by \Alfven waves \citep{hfl,cm95}\footnote{Although these 
works focus on more massive RGB stars than ours, the result will not change 
if less massive stars are considered. } 
The main problem is that a sizable amount of the input wave energy remains in 
the supersonic region, which leads to faster wind. 
There is also a similar problem in the concept of 
a supersonic transition locus, which \citet{mul78} introduced to explain 
the onset of massive wind. 
The main difference of our 
results from the previous works is that the large subsonic regions are formed 
in our simulations due to the nearly static regions above 
the surfaces. Therefore, most of the input \Alfven wave energies are 
deposited in the subsonic regions, which results in the slow winds, 
although the dissipation is not much faster 
than in the previous model calculations.    

We speculate that our dynamical wind solutions 
correspond to ``inner solutions'' introduced by \citet{hfl}. 
They pointed out that there are two transonic solutions in some cases; 
one is the inner solution in which the wind is accelerated very 
gradually and the wind speed is slower, although 
they mainly studied the other one, the outer solution, in which the 
acceleration starts just above the surface. 
Our dynamical simulations automatically select stable transonic wind 
solutions. Then, our results might show that 
the inner solution is the stable branch of \Alfven wave-driven winds. 

We would like to remark on an issue of time-dependent winds. 
Our simulations show that some portions of the RGB atmospheres transiently 
accrete ($v_r<0$), which lasts longer than the assumed longest wave periods 
({\it e.g.} see the dashed line in the top panel of Figure \ref{fig:mpw}). 
This phenomenon may be an example of the hysterisis 
transition between subsonic accretion and supersonic wind discovered 
by \citet{vel94,dvl98}.

\subsection{Limitations}
\label{sec:lim}
Here we discuss processes and effects which need to be studied in more 
detailed calculations. 

\subsubsection{Heating of Low Chromosphere: Sound Waves?}
We do not input longitudinal fluctuations at the photospheres, which would 
directly drive sound waves, because they do not contribute to the heating and 
acceleration of the stellar winds ({\it e.g.} Judge \& Carpenter 1998) 
(In our simulations sound waves are generated from \Alfven waves 
in the upper regions  
; \S \ref{sec:dwv}).  
However, the sound waves directly from the surface might be 
important in the heating of low chromosphere (e.g. 
Carlsson \& Stein 1997). 
We switch off the radiative cooling in the inner regions 
if $T< 3000$ K and $\rho > 5\times 10^{-17}$ g cm$^{-3}$ by the 
technical reason (\S\ref{sec:siml}). 
By taking into account the longitudinal perturbations at the photospheres, 
we may carry out a more self-consistent treatment of the low chromospheres 
without switching off the radiative cooling artificially.

\subsubsection{Multi-dimensional Effects -Wave Generation-}
\label{sec:clp}
In our simulations we only consider the wave generation at the photospheres 
in the 1D flux tubes. However, interactions between flux tubes may be 
important in generation of waves, though 2D or 3D modeling is required to 
study such processes.   
For example, reconnection events between 
closed loops could be sources of small flare-like events, and these events 
probably excite compressive waves at locations above the photosphere 
\citep{str99}. 

The magnetic reconnection between a closed loop and 
an open field line also possibly excites \Alfven waves traveling upwardly
\citep{miy05}. These waves would also play a role in the acceleration 
of red giant winds. Further detailed analysis is necessary to estimate how   
such waves produced by activities involving closed loops are important 
with respect to the energetics.  

\subsubsection{Multi-dimensional Effects -Wave Propagation-}
We treat the propagation of \Alfven waves in the 1D flux tubes. In other  
words, \Alfven waves are assumed to travel in the same way in neighboring 
flux tubes. 
However, if \Alfven speeds are slightly different in neighboring 
field lines, the waves become out of phase even if they are generated 
in the same way at the surface. Then, these \Alfven waves dissipate 
by phase mixing through resistivity and viscosity \citep{hp83}. 

\Alfven waves in stellar winds become more or less 
turbulent-like. Then, \Alfvenic turbulence 
cascades to higher frequency in the transverse direction with respect to 
underlying magnetic fields \citep{gs95}. 
Therefore, \Alfven waves might dissipate by different channels in more 
realistic 3D calculations, while in our 1D simulations the \Alfven waves 
dissipate mainly through the decay instability (\S\ref{sec:dwv}). 

\subsubsection{Collisionless Effects}
We assume the MHD approximation, namely a mean free path with respect to 
Coulomb collisions, 
$l_{\rm mfp} = 9.38\times 10^7\;{\rm cm}
\frac{(T/10^6\;{\rm K})^2}{n/10^8\;{\rm cm^{-3}}},$
is smaller than a typical scale. 
Here, we can take the minimum wavelength, 
$\lambda_{\rm min}\simeq v_{\rm A}\tau_{\rm min}$, of the \Alfven waves 
as a typical scale, hence,  
\begin{equation}
\lambda_{\rm min} > l_{\rm mfp}
\end{equation}
is the requirement for the MHD condition. Otherwise we have to consider 
collisionless effects of plasma. 
This condition breaks in $r\gtrsim 2\;R_{\star}$ of the MS star (Model I) 
and in $\gtrsim 5\;R_{\star}$ of the sub-giant star (Model II), 
while in the RGB stars (Models III - VI) it holds in the entire simulation 
regions because the densities are high and the temperatures are low.  
 
In the collisionless regime, compressive (slow/fast MHD) waves 
suffer collisionless damping 
\citep{bar66, shl06}.
Then, the dissipation rate of the slow MHD waves, which are nonlinearly 
generated from the \Alfven waves, may be changed. This might further 
modify the mode conversion rate from the \Alfven waves. 
Then, the dissipation of the \Alfven waves might be indirectly affected 
in Models I and II.  

\subsubsection{Atoms, Molecules, and Dusts}
\label{sec:pion}
In the RGB atmospheres, the temperatures sometimes become $10^3 - 
10^4$ K as shown in the snap-shot results (Figures \ref{fig:mpw} \&  
\ref{fig:tdpd1M}).
Although even in such low temperatures a fraction of the gas is expected to be 
ionized by the UV/X-ray emissions from hot bubbles, a sizable amount of 
neutral atoms remain, or even molecules can also exist if $T\lesssim 3000$ K 
\citep{tsu73}.  The snap-shot results also show that the temperatures 
partly decreases
further to $\sim 1000$ K, and in such circumstances dusts are also formed. 

The existence of atoms and molecules indicate that the gas is not fully 
ionized but partially ionized. The MHD approximation implicitly assumes 
that the coupling between ions and neutrals is sufficient so that 
the gas can be treated as one fluid. Otherwise, the friction between 
these two components plays a role in the dissipation of \Alfven waves. 
We can check the condition of ion-neutral coupling by comparing wave 
frequency and a ion-neutral collision rate \citep{drd83}. 
We have found that the coupling is sufficient with respect to the 
low-frequency waves we are dealing with; we do not have to 
worry about the effect of weakly ionized gas.  
However, if high frequency \Alfven waves are generated by turbulent 
cascade, these waves dissipate by the ion-neutral damping more effectively 
than in fully-ionized gas \citep{hfl}.

Molecules can be used as diagnostics to understand physical conditions of 
RGB star atmospheres. Observations of CO show that quasi-static cool ($T\sim 
1000 -2000$ K) regions containing molecules are in the atmospheres 
of RGB stars, typically at $\sim 2R_{\star}$ \citep{tsu88, tsu97}.  
Our simulations of the RGB stars also show the quasi-static regions 
in the atmospheres and the temperatures partly becomes low enough to form 
molecules. The cooling by molecules drive thermal instability 
at $T=2000 - 3500$ K \citep{mns87}, 
similarly to the ions in $T\gtrsim 10^5$ K.  
This will be important in controlling thermal properties of the atmospheres 
and winds of RGB stars. 

Radiation pressure on dusts is not efficient in driving winds of RGB 
stars \citep{js91,vj06}, although it could be important in asymptotic giant 
branch stars. 
However, \Alfven waves may dissipate by the cyclotron 
resonance with charged dusts \citep{vj06}, which might also 
give a correction to the dissipation of the \Alfven waves. 
To study this effect, we have to carefully consider the formation processes of 
dusts in intermittently existing cool ($T\sim 1000$ K) gas.

\section{Summary}
We have studied the winds of the intermediate and low mass RGB stars near 
the DL, by comparing with the coronal winds from the MS and 
SB stars. These stars all have the surface convective layers, which play a 
major role in driving the stellar winds as well as the UV/X-ray 
activities because other mechanisms such as centrifugal force and radiation 
pressure are insufficient.  
The surface turbulences excite various types of waves, among which the \Alfven 
wave travels a longer distance and contribute to not only the atmospheric 
heating near the surface but the wind acceleration in $\gtrsim$ 
several $R_{\star}$. 

We have, for the first time, successfully carried out the time-dependent 
MHD simulations of the red giant winds in the open magnetic field regions from 
the photospheres to the sufficiently distant locations 
($\approx 25\; R_{\star}$).   
We inject the fluctuations from the photospheres, of which the amplitudes 
and spectra are estimated from the properties of the surface convection.  
We have determined the velocities, densities, and temperatures of the stellar 
winds in a self-consistent manner by taking into account the radiative 
cooling and thermal conduction; the propagation and dissipation of the MHD 
waves are automatically treated, and the subsequent heating and acceleration 
of the gas are also self-consistently handled by the jump conditions of 
the MHD shocks, using the nonlinear MHD Godunov-MOCCT algorithm.  

The stellar winds are driven by the nonlinear dissipation 
of the \Alfven waves through 
the conversion to compressive waves, whereas other dissipation processes might 
become important when performing 3D modeling/simulations. 
The energy fraction that is transferred to the stellar wind against 
the initial wave energy is quite small, $\sim$ 1\%, and the rest of the 
energy is lost by the wave reflection (less evolved stars) and radiation 
(more evolved stars). However, this tiny fraction is enough to explain the 
observed stellar winds from MS and RGB stars.   

When the stars evolve to slightly blueward positions of the DL, 
the steady hot coronae with temperature, 
$T \gtrsim 10^6$ K, disappear. 
In the atmospheres and winds of the RGB stars magnetized hot bubbles and cool 
chromospheric gas co-exist. The hot bubbles are created in fast MHD shocks and 
supported by the magnetic pressure. 
Therefore, the densities of the bubbles can be 
kept low.  The radiative cooling is reduced to balance with the heating so 
that the bubbles survive much longer than the cooling timescale.  

The red giant winds are structured with many bubbles and blobs, and the mass 
loss rate and wind speed show large time-variations.     
The hot bubbles possibly become sources of EUV and soft X-ray emissions 
of hybrid stars, whereas the cool chromospheric wind might absorb the 
emissions. 
The temperatures of the hot bubbles become lower in more evolved and 
less massive stars, which is consistent with the trends inferred from the 
observations of hybrid stars.

The wind velocities also drop 
to 10-100 km s$^{-1}$ in the RGB stars, which are much slower than the 
escape speeds at the stellar surfaces.  This is mainly because the static 
atmospheres are formed above the photospheres and the stellar winds are 
practically accelerated from several stellar radii. 
The terminal velocities are regulated by the escape speeds there, and 
the slower winds can escape. 

We have finally derived a relation (Equation \ref{eq:mswv}) that determines 
the mass loss rates of MS to RGB stars from the energetics of the 
simulated wave-drive 
winds. This has a different dependence on the stellar parameters from 
the well known Reimers' formula which mainly focuses on more evolved giants. 
The derived relation gives a better explanation of $\dot{M}$ of less luminous 
stars and can be used complimentarily to the Reimers' formula.    
 

The author thanks Profs. T. Tsuji, H. Shibahashi, H. Ando, T. Watanabe, and 
S. Inutsuka for many fruitful discussions. The author also thanks the referee 
for many constructive comments. 
This work is supported in part by a Grant-in-Aid for Scientific
Research (18840009) from the Ministry of
Education, Culture, Sports, Science, and Technology of Japan.

\end{document}